\documentclass[%
superscriptaddress,
reprint,
showpacs,
 amsmath,amssymb,
 aps,
prd,
showkeys
]{revtex4-1}

\usepackage{graphicx}
\usepackage{dcolumn}
\usepackage{bm}
\usepackage[bookmarks=true,bookmarksopen=true, bookmarksnumbered=true]{hyperref}
\usepackage[mathlines]{lineno}
\usepackage{amsfonts, amsmath, amssymb}
\usepackage{xcolor}
\usepackage{siunitx}
\usepackage[list=true, 
labelformat=brace]{subcaption}
\usepackage{float}
\usepackage{multirow}

\begin{document}


\title{Entropy methods for CMB analysis of anisotropy and non-Gaussianity}

\author{Momchil Minkov}
\email{mminkov@stanford.edu}
\affiliation{Department of Electrical Engineering, Stanford University, 350 Serra Mall, Stanford, CA 94305-9505, USA}

\author{Marvin Pinkwart}
\email{m.pinkwart@jacobs-university.de}
\affiliation{Department of Physics and Earth Sciences, Jacobs University Bremen, Campus Ring 1, 28759 Bremen, Germany}
 
\author{Peter Schupp}%
 \email{p.schupp@jacobs-university.de}
 \affiliation{Department of Physics and Earth Sciences, Jacobs University Bremen, Campus Ring 1, 28759 Bremen, Germany}

\date{\today}

\begin{abstract}
In recent years, high-resolution cosmic microwave background (CMB) measurements have opened up
the possibility to explore statistical features of the temperature fluctuations down to very small angular
scales. One method that has been used is the Wehrl entropy, which is, however, extremely costly in terms of
computational time. Here, we propose several different pseudoentropy measures (projection, angular, and
quadratic) that agree well with the Wehrl entropy, but are significantly faster to compute. All of the
presented alternatives are rotationally invariant measures of entanglement after identifying each multipole $l$
of temperature fluctuations with a spin-$l$ quantum state and are very sensitive to non-Gaussianity,
anisotropy, and statistical dependence of spherical harmonic coefficients in the data. We provide a simple
proof that the projection pseudoentropy converges to the Wehrl entropy with increasing dimensionality of
the ancilla projection space. Furthermore, for $l=2$, we show that both the Wehrl entropy and the angular
pseudoentropy can be expressed as one-dimensional functions of the squared chordal distance of multipole
vectors, giving a tight connection between the two measures. We also show that the angular pseudoentropy
can clearly distinguish between Gaussian and non-Gaussian temperature fluctuations at large multipoles
and henceforth provides a non-brute-force method for identifying non-Gaussianities. This allows us to
study possible hints of statistical anisotropy and non-Gaussianity in the CMB up to multipole $l=1000$
using \textit{Planck} 2015, \textit{Planck} 2018, and \textit{\textit{WMAP}} 7-yr full sky data. We find that $l=5$ and $l=28$ have a large
entropy at $2$--$3\sigma$ significance and a slight hint towards a connection of this with the cosmic dipole. On a
wider range of large angular scales we do not find indications of violation of isotropy or Gaussianity. We
also find a small-scale range, $l\in[895,905]$, that is incompatible with the assumptions at about $3\sigma$ level,
although how much this significance can be reduced by taking into account the selection effect, i.e.,, how
likely it is to find ranges of a certain size with the observed features, and inhomogeneous noise is left as an
open question. Furthermore, we find overall similar results in our analysis of the 2015 and the 2018 data.
Finally, we also demonstrate how a range of angular momenta can be studied with the range angular
pseudoentropy, which measures averages and correlations of different multipoles. Our main purpose in this
work is to introduce the methods, analyze their mathematical background, and demonstrate their usage for
providing researchers in this field with an additional tool. We believe that the formalism developed here can
underpin future studies of the Gaussianity and isotropy of the CMB and help to identify deviations,
especially at small angular scales.

\end{abstract}

\maketitle


\section{Introduction}

Since its discovery in 1964 by Penzias and Wilson \cite{Penzias}, the cosmic microwave background (CMB) has served as the main source of information about the current and past Universe. Originating in the process of recombination at about 380$\,$000 years after the Big Bang at redshift of about 1100, the CMB temperature distribution on the celestial sphere displays the energy density distribution on the Last Scattering Surface when the Universe became transparent to electromagnetic radiation. The CMB intensity follows a nearly perfect Planckian distribution and its average temperature has been measured to be $T_0=2.72548 \pm 0.00057 \, \mathrm{K}$\cite{Fixsen} with minor fluctuations of order $10^ {-5}$--$10^{-4} \, \mathrm{K}$ and one larger dipole modulation of order $10^{-3}\, \mathrm{K}$, which is assumed to be of pure kinematic origin. The anisotropy of the CMB temperature has been measured first by the \textit{Cosmic Background Explorer} satellite from 1989 to 1993, followed up by the Wilkinson Microwave Anisotropy Probe (\textit{\textit{WMAP}}) mission from 2001 to 2010. The most recent CMB investigation satellite, \textit{Planck}, was launched in 2009 and shutdown in 2013. Its 2015 results from the second data release provide the most precise values of cosmological parameters \cite{Planck2015I,Planck2015XIII} measured up to this point. Recently, they have been refined in the 2018 data release \cite{Planck2018I,Planck2018VI}.

It is commonly assumed that the tiny temperature fluctuations follow a Gaussian and statistically isotropic distribution. This assumption has been confirmed by the \textit{Planck} mission to a large extent\cite{Planck2015XVI,Planck2015XVII}, but nevertheless the search for possible non-Gaussianities\cite{KomatsuNonGaussianity} and statistical anisotropies \cite{Schwarz2004,schupp1,Abramo,Schwarz2007,Aurich,Schwarz2009,Gruppuso,Schwarz2010,Zunckel,Sung} has been rich and certain anomalies have been found, as, for example, unusual (anti-)correlation of the lowest multipoles with the Cosmic Dipole as well as with each other, a sign of parity asymmetry and a lack of large-angle correlation (see e.g. the review\cite{Schwarz2015}). 

Common tools in these analyses are multipole vectors (MPVs) which were introduced for cosmological data analysis in \cite{MultiCopi} and whose properties have been elaborated in \cite{schupp1,KatzWeeks,Dennis2004,Dennis2005,Dennis2007}. For the most recent results on possible CMB anomalies using multipole vectors and an overview over the mathematical approaches see \cite{Pinkwart,PinkwartSchwarz,Oliveira}.

MPVs are closely related to Bloch coherent states (see \cite{schupp1}) which were also used in the past to prove special cases of Lieb's conjecture\cite{Schupp} for the Wehrl entropy. 

In this work, we develop and compare several rotationally invariant measures of randomness on functions on the two-sphere, namely, the angular, projection, and quadratic pseudoentropies. We show that for $l=2$ the Wehrl and angular entropy can be expressed as a function of the squared chordal distance of MPVs. We find that all these measures except the quadratic one show the same features, making the quadratic pseudoentropy the least preferred measure. Because of the shared features, we then restrict ourselves to the numerically fastest method, and use it to analyze \textit{Planck} 2015 and 2018 full sky as well as \textit{WMAP} 7-year Internal Linear Combination (ILC) maps. The angular pseudoentropy allows for comparing the data to many ensembles of Gaussian and isotropic random maps up to $l=1000$ in short computing time. With a better theoretical understanding of confidence levels, also the Wehrl entropy could be used easily since the computing time for a single map is still reasonable. In general, it is especially nice to have a single number for each multipole even in the case that the data would not be Gaussian and isotropic. In this case, the CMB would be described by more than one degree of freedom (d.o.f.) per multipole. Non-Gaussian distributions need higher correlation functions and anisotropic distributions yield an $m$-dependent two-point function. In the tradition of thermodynamics, with these pseudoentropies one can approximately reduce a possibly large set of data again to one number for each multipole. Since all considered types of entropies show a similar behavior the information does not depend on the definition of the entropy. Eventually there exists also an extension of the angular entropy to ranges and collections of multipoles, which we call range angular entropy.

This paper is organized as follows: In Sec.~\ref{CMB statistics}, we briefly recapitulate the basic ingredients of CMB spherical harmonic statistics. Afterwards, in Sec.~\ref{math}, we introduce our methods mathematically, clarify their properties, and show the connection to multipole vectors. We also provide a simple proof of the convergence of the projection entropy to the Wehrl entropy up to a term which is independent of the input density matrix. Section \ref{Results} is dedicated to the application of our methods to real data. We compare the different pseudoentropy methods, then we apply the angular pseudoentropy to 2015 \textit{Planck} and 7-year \textit{WMAP} full sky foreground-cleaned maps before comparing the 2015 results to those obtained with 2018 data and also applying the range entropy and comparing it to the statistics used before. Eventually, in Sec.~\ref{sum} we summarize and discuss our findings.

\section{CMB statistics}
\label{CMB statistics}
As a function on $S^2$ the CMB temperature fluctuations $\Delta T := \delta T/T_0$ can be decomposed uniquely according to irreducible representations of $SO(3)$, i.e.,, into spherical harmonics
\begin{equation}
\Delta T(\theta,\phi) = \sum_{l=1}^{\infty}\sum_{m=-l}^l a_{lm}Y_{lm}(\theta,\phi) \, \in \mathbb{R},
\end{equation}
where the $l=0$-summand is omitted because the fluctuations average to zero, and $\theta$, $\phi$ denote the usual spherical coordinates. The fact that $\Delta T$ is real together with the property $Y_{l,-m} = (-1)^m Y_{lm}^* $ impose the constraints
\begin{equation}
a_{l,-m} = (-1)^m a_{lm}^*
\end{equation}
on the spherical harmonic coefficients, leaving for each multipole number $l$ exactly $2l+1$ real d.o.f. The multipole number $l$ corresponds to angular scales of $\approx 180/l\, \deg$. The orthonormality of $\{Y_{lm}\}$ allows to compute the coefficients from the temperature map via
\begin{equation}
a_{lm} = \int_{S^2}\! \mathrm{d}\Omega \, \Delta T(\Omega)Y_{lm}^*(\Omega).
\end{equation}
Simple inflationary models together with linear perturbation theory predict nearly Gaussian temperature fluctuations and a further common assumption is statistical isotropy. The spherical harmonic coefficients inherit both properties from the temperature map, meaning that 
\begin{equation}
p(\vec{a}_l) = \frac{1}{\mathcal{N}}e^{-\frac{1}{2}\vec{a}_l^{\dagger} \mathbf{D}_l \vec{a}_l} \quad \text{(Gaussianity),}
\end{equation}
where $p$ denotes the joint probability distribution, $\vec{a}_l = (a_{l0},\ldots,a_{ll})^T$, $C^l_{mn} = \left( \mathbf{D}_l^{-1} \right) _{mn} = \langle a_{lm}a^*_{ln}\rangle$ and $\mathcal{N}$ denotes a normalization constant, and
\begin{align}
\forall \mathbf{R} \in SO(3), \vec{e}_1,\ldots,\vec{e}_n \in S^2, n \in \mathbb{N}: \nonumber \\
G_n(\{\mathbf{R}\vec{e}_i\}) = G_n(\{\vec{e}_i\}) \quad \text{(isotropy),}
\end{align}
where $G_n(\{\vec{e}_i\}) = \langle \prod_{i=1}^n \Delta T (\vec{e}_i) \rangle$ denotes the $n$-point function of temperature fluctuations. The isotropy condition is equivalent to rotationally invariance of the joint $a_{lm}$-probability distribution. The averaging $\langle . \rangle$ is meant to be performed over all possible universes, which of course is not possible, wherefore we are left with a natural and inevitable variance in all quantities, called cosmic variance. If we impose both Gaussianity and isotropy then the two-point correlation of spherical harmonic coefficients is diagonal
\begin{equation}
\label{cl_eq}
C^l_{mn} = C_l \delta_{mn}.
\end{equation}
In practice, the power spectrum $C_l$ is calculated using the unbiased estimator
\begin{equation}
\label{cl_estimator}
\hat{C}_l = \frac{1}{2l+1}\sum_m |a_{lm}^2| \, , \, \langle \hat{C}_l \rangle = C_l,
\end{equation}
with cosmological variance
\begin{equation}
\mathrm{var}(\hat{C}_l) = \frac{2}{2l+1}C_l^2.
\end{equation}

\section{pseudoentropies and their properties}
\label{math}

In this section, we shall discuss the mathematical properties and interrelation of various macroscopic entropy measures that can be used as powerful tools to analyze Gaussianity and isotropy of the CMB and can also be useful in other contexts. This section contains a review of the mathematical background as well as new definitions, results and insights. The motivation to look for macroscopic entropy measures is the same as in statistical physics: A microscopic description of a physical system, e.g., the positions and momenta of a fluid, is useful for simulation purposes, but not when comparing to a real fluid. Instead, one would resort to the study of macroscopic quantities and parameters like internal energy, temperature, entropy, pressure etc.~that are well-defined because of symmetries. In the analysis of the CMB, the $a_{lm}$ coefficients are an analog of the microscopic quantities. For low $l$ these and derived quantities like multipole vectors can be studied individually, but for high~$l$ this quickly becomes impractical: The \textit{Planck} mission data easily comprises several million reliable data points. For the CMB the obvious underlying spacetime symmetry is rotation invariance. The representations of the rotation group decompose into irreducible components labeled by the angular momentum quantum number $l$ (multipole expansion) and we can focus on fixed-$l$ subspaces. A loose analog of internal energy is the angular power spectrum, i.e., the $C_l$ coefficients [see Eqs.~\eqref{cl_eq} and \eqref{cl_estimator}]. They have proven immensely useful in the analysis of the CMB and its cosmological implications, but when it comes to questions of isotropy and preferred directions, individual $m$ matter and an analog of entropy would be useful. A natural idea is to consider the abstract quantum state $|\psi\rangle := \sum a_{lm} |l,m\rangle$ that can be formally computed from the $a_{lm}$ and associate an entropy $S$ to it. We will usually focus on one $l$ at a time and normalize the states by rescaling the~$a_{lm}$ appropriately. Since the states are by construction pure, the von Neumann entropy will be trivially zero, but there are also non-trivial pseudoentropies that can distinguish pure states and turn out to be sensitive to non-Gaussianity and anisotropy. The general strategy is as follows:
\begin{equation}
\begin{split}
T(\theta,\phi) &\rightarrow a_{lm} \rightarrow |\psi\rangle := \sum a_{lm} |l,m\rangle \\
& \rightarrow \rho = |\psi\rangle\!\langle \psi| \rightarrow \rho_\text{mixed} \rightarrow S \ ,
\end{split}
\end{equation}
where $\rho_\text{mixed}$ is obtained from the pure state $\rho$ by applying a rotationally symmetric quantum channel $\Phi$, i.e., a completely positive map between Hilbert spaces with possibly different dimensions, or by computing its lower symbol, i.e., its expectation value in spin coherent states. The latter choice leads to the Wehrl entropy \cite{Wehrl2,Lieb78,Schupp}
\begin{equation}
S_W = - (2l+1) \int \frac{d\Omega}{4\pi} |\langle \Omega|\psi\rangle|^2 \ln |\langle \Omega|\psi\rangle|^2 \ , \label{Wehrlentropie}
\end{equation}
where $|\Omega\rangle$ is a spin-$l$ coherent state. The Wehrl entropy was first proposed in \cite{Helling} as a useful tool for CMB analysis. See Fig.~\ref{wehrl_30} for a  showing the Wehrl entropy for CMB data. Closely related  is the ``quadratic entropy'' that is obtained by replacing $-x \ln x$ in the formula for the Wehrl entropy by the concave function $x(1-x)$,
\begin{equation}
S_{\text{quad}} = 1 - \frac{2l+1}{4l+1} \left|P_{2l} |\psi \otimes \psi \rangle\right|^2 \ , \label{quadraticentropie}
\end{equation}
where $P_{2l}$ is the projector onto the spin-$2l$ part, i.e., the highest spin component of the tensor product.
Other examples using the choice $-x \ln x$ are what we call angular entropy
\begin{align}
S_{\text{ang}} =& \, \mathrm{Tr} \left[ \phi\Big(\sum_{i=1}^3 \frac{L_i |\psi\rangle\!\langle \psi| L_i}{l(l+1)}  \Big) \right] \\
\label{phi_def}
\text{with} \quad \phi(x) :=& -x \ln x \ ,
\end{align}
where the $L_i$ are angular momentum generators in the spin-$l$ representation, and $j$-projection entropy
\begin{equation}
S_{\text{proj}}^{(j)} =  \mathrm{Tr} \left[\phi\Big(\frac {2l+1}{2(l+j)+1} P_{l+j}\big(|\psi\rangle\!\langle \psi| \otimes \mathbf 1\big)P_{l+j}\Big) \right]
\end{equation}
where $\mathbf 1$ is the unit operator on a spin-$j$ ancilla $[j]$ and $\phi$ is as in Eq.~\eqref{phi_def}.
An overview of these entropies applied to CMB data can be found in Fig.~\ref{Ent_comp}. In the following section we shall explain the mathematics in detail, derive relations between the various entropies and point out interesting side results including a fast way to compute multipole vectors. Readers that are mostly interested in results and numerics can skip to the algorithm \eqref{ang_alt}-\eqref{ang_range} at the end of Sec.~\ref{Math_a}.

\subsection{Coherent states, multipole vectors and entropy}
\label{Math_a}

Coherent states were originally introduced by Schr\"{o}dinger \cite{Schrodinger} and are well known in the context of the quantum harmonic oscillator, where they can be defined either as eigenstates of the lowering operator or, equivalently, as elements of the orbit of the ground state under the Heisenberg group. Perelomov \cite{Perelomov} has generalized the latter notion to orbits of a fiducial vector in some representation of a Lie group under the action of that group. The choice of the fiducial vector is essential for the properties of the resulting coherent states. 
Spin coherent states -- also called Bloch coherent states -- in a spin-$l$ irreducible representation $[l] \equiv {\mathbb C}^{2l+1}$ of $SU(2)$ with $2l+1 \in \mathbb N$
are defined as orbits of the highest weight vector $|l,l\rangle$. The stability group of that vector  is $U(1)$ and  spin coherent states can thus be labeled by points $\Omega =(\theta, \phi)$ on the sphere $S_2 \cong SU(2)/U(1)$,
\begin{equation}
|\Omega_l\rangle = \mathcal R(\Omega) |l,l\rangle \ ,
\end{equation}
where $\mathcal R(\Omega)$ denotes a rotation that takes the north pole to the point $\Omega$ and $l$ labels the representation of $SU(2)$. For the CMB data $l$ will be an integer, but everything we discuss here is also valid for half-integer $l$.
For $l= \frac 12$ this gives for example
\begin{equation}
|\Omega_{\frac 12}\rangle = e^{-i\phi/2} \cos \tfrac\theta 2 |\tfrac 12, \tfrac 12\rangle
+ e^{+i\phi/2} \sin \tfrac\theta 2 |\tfrac 12, -\tfrac 12\rangle \ . \label{spin12}
\end{equation}
Coherent states inherit nice properties from the underlying fiducial vector. A particular important one is that the tensor product of coherent states is again a coherent state and lies in the highest spin component:
\begin{equation}
|\Omega_l\rangle \otimes |\Omega_{j}\rangle 
=  |\Omega_{l+j}\rangle \label{productproperty}
\end{equation}
Using this property repeatedly yields an explicit formula for any spin from \eqref{spin12}:
\begin{equation}
\label{factorcoherent}
\begin{split}
|\Omega_l\rangle &= |\Omega_{\frac 12}\rangle \otimes \ldots \otimes |\Omega_{\frac 12} \rangle \\
&= \sum_{m=-l}^l \binom{2l}{l+m}^{\!\frac 12} e^{-im \phi /2}\,\cos^{l+m}(\tfrac\theta 2) \, \sin^{l-m}(\tfrac\theta 2) \, |l,m\rangle
\end{split}
\end{equation}
Interestingly, such a product representation in terms of spin-$\tfrac 12$ states exists for any state  $|\psi\rangle \in [l]$, but except for coherent states, a projection onto the highest spin component and a renormalization are required \cite{Schupp}:
\begin{equation}
|\psi_l\rangle = \sum_{m=-l}^l  a_{lm} |l,m\rangle = c P_l\Big( |\Omega^{(1)}_{\frac 12}\rangle \otimes \ldots \otimes |\Omega^{(2l)}_{\frac 12} \rangle\Big) \ , \label{factorpsi}
\end{equation}
where $P_l$ is the projector onto $[l]$ and  $c$ is a normalization constant. 
The~$\Omega^{(i)}$ point into the direction of the $2l$ multipole vectors that characterize the state~$|\psi\rangle$. Contracting \eqref{factorcoherent} with \eqref{factorpsi} and using the stereographic projection to express points on the sphere in terms of complex numbers $z = e^{i\phi}\cot (\tfrac \theta 2)$, leads to a polynomial 
\begin{equation}
\sum_{m=-l}^l \binom{2l}{l+m}^{\!\frac 12} z^{l+m} a_{lm} \ ,
\end{equation}
whose $n \leq 2l$ zeros (roots) correspond to points on the sphere that are antipodal to $n$ of the $2l$ multipole vectors. The remaining $2l -n$ multipole vectors point to the south pole of the sphere. For the CMB data $l$ is an integer, $\Delta T(\Omega)$ is real and consequently $a^*_{lm} = (-)^m a_{l,-m}$. This implies that the zeroes of the polynomial are located at pairs of antipodal points on the sphere and the multipole vectors come in anti-aligned pairs (for details see \cite{Helling}). In \cite{Schupp} this method was introduced to determine explicit formulas for the Wehrl entropy and to prove Lieb's conjecture. Applying those explicit formulas to the case $l = 2$ with two pairs of anti-aligned multipole vectors of length $1/2$ gives the following formula for the Wehrl entropy as a function of the squared chordal distance $\epsilon  = \sin^2 (\frac{\alpha}2)$ between the vectors, where $\alpha$ is the angle between them:
\begin{equation}
S_{\mathrm{W}}(\epsilon ) = c - \ln c + \frac{32}{15} - \ln 6 \ ,
\end{equation}
where
\begin{equation}
c = c(\epsilon) := \frac1 {1 -\epsilon (1-\epsilon )} \ .
\end{equation}
For the angular entropy a similar computation gives 
\begin{equation}
\label{Sang(eps)}
S_\text{ang}(\epsilon ) =- \frac c2 \left((1-\epsilon )^2 \ln (1-\epsilon )^2 + \epsilon ^2 \ln \epsilon ^2\right) - \ln \frac c2 \ 
\end{equation}
with $c$ as above.
\begin{figure}
\begin{center}
\includegraphics[width=0.5\textwidth]{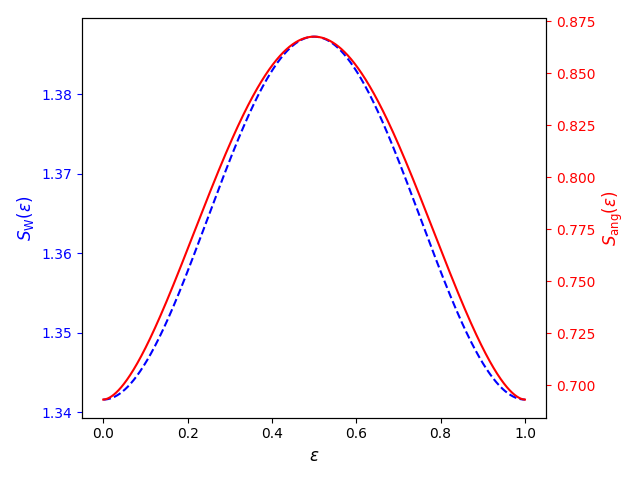}
\caption{Dependence of the angular pseudoentropy $S_{\mathrm{ang}}(\epsilon)$ (solid) and the Wehrl entropy $S_{\mathrm{W}}(\epsilon)$ (dashed) on the squared chordal distance $\epsilon$ between multipole vectors on the sphere with radius $r=\frac{1}{2}$ for $l=2$. The maximum is obtained when the
multipole vectors are orthogonal to each other and the minimum
is obtained when both multipole vectors are the same.
}
\label{BothS}
\end{center}
\end{figure}
Plots of the two functions look very similar (see Fig.~\ref{BothS}), confirming the observed similarities in behavior of the two entropy measures in the CMB analysis (see Fig.~\ref{Ent_comp}).
The polynomial method provides a very fast and convenient way to determine multipole vectors and has been used in \cite{Helling,Dennis2007} and many other publications to analyze the CMB.

Spin coherent states are complete via Schur's lemma
\begin{equation}
(2l +1) \int \frac{d \Omega}{4\pi}\, |\Omega_l\rangle\!\langle\Omega_l| =  P_l \ , \label{complete}
\end{equation}
where $P_l$ is the projector onto $[l]$. They are normalized $\langle\Omega_l|\Omega_l\rangle = 1$ but not orthogonal
\begin{equation}
|\langle\Omega_l | \Omega'_l\rangle|^2 = \cos^{4l} (\mbox{\large$\sphericalangle$}{(\Omega, \Omega')}) \ , \label{notorthogonal}
\end{equation}
i.e.,\ they form an overcomplete basis of $[l]$. 
In the $l \rightarrow \infty$ limit, $(2l+1) |\langle \Omega_l | \Omega'_l\rangle|^2$ becomes a delta function $\delta(\Omega, \Omega')$  and in this limit the coherent states form an infinite-dimensional orthonormal basis labeled by points on the sphere.

A striking property of coherent states is that the diagonal matrix elements 
\begin{equation}
A(\Omega) = \langle\Omega_l|A|\Omega_l\rangle \quad  \text{ (lower symbol)} 
\end{equation}
of an operator $A$ on $[l]$ already determine that operator uniquely: Let $C = A - B$ with an arbitrary operator $B$, then $C(\Omega) = 0$ for all $\Omega$ implies $C = 0$, i.e., $A=B$. The proof uses analytic properties of the lower symbol.  The lower symbol is thus a faithful representation of an operator. 
Using Eq.~\eqref{complete}, the trace of an operator $A$ on $[l]$ can be computed as an integral over its lower symbol
\begin{equation}
\mathrm{Tr}_{[l]} (A) =  (2l+1)  \int \frac{d \Omega}{4\pi} \,   \langle\Omega_l|A|\Omega_l\rangle  =  (2l+1)  \int \frac{d \Omega}{4\pi} \,  A(\Omega) \ . \label{traceformula}
\end{equation}
Another interesting property is that any operator $A$ on $[l]$ can be expanded diagonally in coherent states
\begin{equation}
A = (2l+1)  \int \frac{d \Omega}{4\pi} \, h_A(\Omega) |\Omega_l\rangle\!\langle\Omega_l| \ , \label{upper}
\end{equation}
where $h_A(\Omega)$ is called an upper symbol of $A$. These two properties are in fact closely related: Contracting Eq.~\eqref{upper} with an operator $C$ gives
$\mathrm{Tr} (C^\dagger A) \propto \int d \Omega \, h_A(\Omega)\, \Omega^*(C)$, i.e., the operators that can be represented by an upper symbol as in Eq.~\eqref{upper}, are orthogonal to the operators that are in the kernel of the lower symbol map.  Hermitean operators have real lower and upper symbols. Positive semi definite operators and density matrices have unique non-negative lower symbols, but the same is in general not true for upper symbols. Following Wehrl, these properties suggest to interpret the lower symbol of a density matrix $\rho$, which is by definition positive semi definite and normalized, as a probability density and compute 
\begin{equation}
(2 l +1) \int \frac{d \Omega}{4\pi} \,  \phi( \rho(\Omega)) = (2 l +1) \int \frac{d \Omega}{4\pi} \, \phi(c) \ .
\end{equation}
With $\phi(x) = x$ we can verify the normalization and get $\mathrm{Tr}( \rho) = 1$. With $\phi = - x\ln x$ we compute the Shannon entropy of the probability density $\rho(\Omega)$, which  is precisely the Wehrl entropy 
\begin{equation}
S_W(\rho) = -(2l +1) \int \frac{d \Omega}{4\pi} \, \rho(\Omega) \ln \rho(\Omega) \ , \label{Wehrl}
\end{equation}
with the special case \eqref{Wehrlentropie} for a pure state $\rho = |\psi\rangle\!\langle\psi|$. The Wehrl entropy was introduced as a semi-classical entropy, which is mathematically better behaved than the Boltzmann entropy in classical statistical mechanics. The Wehrl entropy is always larger than the von Neumann entropy and it is positive even for pure states. Furthermore its definition is rotationally symmetric and it is hence perfectly suited for our purposes. The minimum of the Wehrl entropy is attained for coherent states. This fact is surprisingly difficult to prove. It was first shown for low spin in \cite{Schupp} and then finally in general in \cite{Lieb14,Lieb16}. Computational evidence suggested in fact an analogous but much stronger conjecture for any concave function $\phi(x)$ \cite{Schupp,Schupp_unpub}, which has also been settled affirmatively in \cite{Lieb16}. For our application, the maximal value of the Wehrl entropy is more interesting. The exact value is not known, in fact finding the maximizing pure state is another hard problem, but a reasonable upper limit can be obtained very simply from a totally mixed state: $S_W \leq \ln(2l +1)$. To summarize: The Wehrl entropy is the Shannon entropy of the probability density obtained from the (faithful) lower symbol representation of a density matrix. It has all the right properties for our purposes. The only drawback is that its computation with suitable precision has a high computational complexity. We will now introduce and discuss several alternatives with similar properties, but better computability.

Instead of $- x\ln x$, one can consider other concave functions defined on the interval~$[0,1]$. For $\phi(x) = x(1-x)$ we obtain the quadratic entropy
\begin{equation}
S_\text{quad}(\rho) = 1 - \frac{2l+1}{4l+1} \mathrm{Tr}\left(P_{2l} (\rho \otimes \rho )\right) \ ,
\end{equation}
where $P_{2l}$ is the projector onto spin $2l$ and we have used the following trick \cite{Schupp}:
\begin{align*}
\left(\rho(\Omega)\right)^2 &= \langle\Omega_l|\rho|\Omega_l\rangle^2 \\ 
&= \langle\Omega_l\otimes \Omega_l|\rho\otimes \rho|\Omega_l\otimes \Omega_l\rangle 
= \langle\Omega_{2l}|\rho\otimes \rho|\Omega_{2l}\rangle 
\end{align*}
and the trace formula \eqref{traceformula} adapted to spin-$2l$. For a pure state $\rho = |\psi\rangle\!\langle\psi|$ we obtain formula \eqref{quadraticentropie}. The quadratic entropy has a large computational advantage as compared to the Wehrl entropy (see Fig.~\ref{Time}), but its qualitative behavior is a bit different, as can be seen in Fig.~\ref{quad_30}. This is not surprising, since many important properties of entropy like additivity and (strong) sub-additivity depend crucially on the choice of the function $\phi(x) =- x\ln x$. We shall hence not pursue quadratic entropy any further and will try to identify and construct other alternatives of the Wehrl entropy that share its characteristic features but are computationally more accessible. Let us start by reconsidering the main ingredient of Wehrl entropy.

Let $\rho$ be a density matrix on  $[l] = \mathbb C^{2l+1}$  and introduce an ancilla Hilbert space $[j] = \mathbb C^{2j+1}$. Using the product property \eqref{productproperty} and normalization of coherent states, we can rewrite the lower symbol $\rho(\Omega)$ that enters the formula for the Wehrl entropy as follows:
\begin{equation}
\begin{split}
\langle\Omega_l|\rho|\Omega_l\rangle &= \langle\Omega_l|\rho|\Omega_l\rangle \langle\Omega_j|\Omega_j\rangle
= \langle\Omega_l \otimes \Omega_j|\rho \otimes \mathbf 1|\Omega_l \otimes \Omega_j\rangle \\
&= \langle\Omega_{l+j}|\rho \otimes \mathbf 1|\Omega_{l+j}\rangle \ ,
\end{split}
\end{equation}
where  $\mathbf 1$ is the unit operator on $[j]$. The values of the lower symbol are thus the diagonal elements of a family of infinite-dimensional matrices
\begin{equation}
\rho_j(\Omega, \Omega') = \langle\Omega_{l+j}|\rho \otimes \mathbf 1|\Omega'_{l+j}\rangle \ . \label{infdimmatrix}
\end{equation}
By an infinite-dimensional compact analog of the Schur-Horn theorem the diagonal elements $\rho(\Omega)$ are majorized by the eigenvalues of the $\rho_j(\Omega, \Omega')$ matrices. This implies that any concave function of the values $\rho(\Omega)$ will be larger or equal to the respective function of the eigenvalues of $\rho_j(\Omega, \Omega')$. The Wehrl entropy is therefore larger than or equal to the von Neumann entropy of $\rho_j(\Omega, \Omega')$. For convex functions the inequalities are reversed. See e.g.~\cite{Bengtsson} for an overview of the mathematical background.  In the limit $j \rightarrow \infty$ and in view of Eq.~\eqref{notorthogonal} the off-diagonal matrix elements of $\rho_j(\Omega, \Omega')$ become zero and the inequalities become equalities. 
Using the property \eqref{complete} on both sides of Eq.~\eqref{infdimmatrix} we can recover a finite-dimensional matrix 
\begin{equation}
P_{l+j}\left(\rho \otimes \mathbf 1\right) P_{l+j} \label{projmatrix}
\end{equation}
from $\rho_j(\Omega, \Omega')$, where $P_{l+j}$ is the projector onto the highest spin component $[l+j]$ of the tensor product. The matrix \eqref{projmatrix} has the same eigenvalues as  $\rho_j(\Omega, \Omega')$. In fact, if
\begin{equation}
P_{l+j}\left(\rho \otimes \mathbf 1\right) P_{l+j} |V_\lambda\rangle = \lambda |V_\lambda\rangle \label{eigenvalproj}
\end{equation}
then $V_\lambda (\Omega) := \langle \Omega_{l+j} | V_\lambda\rangle$ satisfies
\begin{equation}
(2(l+j) + 1) \int \frac{d \Omega'}{4\pi} \,  \langle\Omega_{l+j}|\rho \otimes \mathbf 1|\Omega'_{l+j}\rangle V_\lambda(\Omega') = \lambda V_\lambda(\Omega) \label{eigenvalint}
\end{equation}
and vice versa if $V_\lambda(\Omega)$ is a solution of Eq.~\eqref{eigenvalint}, then $|V_\lambda\rangle = (2(l+j) +1) \int \frac{d \Omega}{4\pi} \, |\Omega_{l+j}\rangle V_\lambda (\Omega)$ satisfies Eq.~\eqref{eigenvalproj}. We have shown that the eigenvalues of the matrix \eqref{projmatrix} majorize the values of the lower symbol of $\rho$ in the sense explained above, namely that inequalities are implied for concave (or convex) functions of these values. It can furthermore be shown that pure states majorize mixed ones and that among the pure states, projectors $|\Omega\rangle\!\langle \Omega|$ onto coherent states will lead to matrices \eqref{eigenvalproj} that majorize all other choices.
Among the concave functionals we are in particular interested in  entropy and define an appropriately normalized mixed density matrix
\begin{equation}
\rho_{\text{proj}}^{(j)} = \frac{2l+1}{2(l+j)+1} P_{l+j}\left(\rho \otimes \mathbf 1\right) P_{l+j} \ , \label{mixproj}
\end{equation}
whose von Neumann entropy is what we call the ``projection entropy'' 
\begin{equation}
S_{\text{proj}}^{(j)}(\rho) = \mathrm{Tr}\left[ \phi\Big(\frac {2l+1}{2(l+j)+1} P_{l+j}\big(\rho\otimes \mathbf 1\big)P_{l+j}\Big) \right],
\end{equation}
with $\phi(x) = x \ln(x)$.
From the fact that the mixed density matrix \eqref{mixproj} has at most $2j+1$ non-zero eigenvalues, we get an upper bound for the projection entropy \cite{Buciumas} $S_{\text{proj}}^{(j)}(\rho)  \leq \ln(2j+1)$.
From the $[l+j]$ perspective the Wehrl entropy should also be computed from Eq.~\eqref{mixproj} and we get the aforementioned inequalities. The only differences from the original definition of Wehrl entropy \eqref{Wehrl} is a rescaling of the density matrix and related renormalization of the integral, which leads to a shift in entropy and the following inequality:
\begin{equation}
\label{proj_wehrl_conv}
S_W(\rho) \geq S_{\text{proj}}^{(j)}(\rho) + \ln\left( \frac {2l+1}{2(l+j)+1}\right) \ .
\end{equation}

\begin{figure}
\begin{center}
\includegraphics[width=0.5\textwidth]{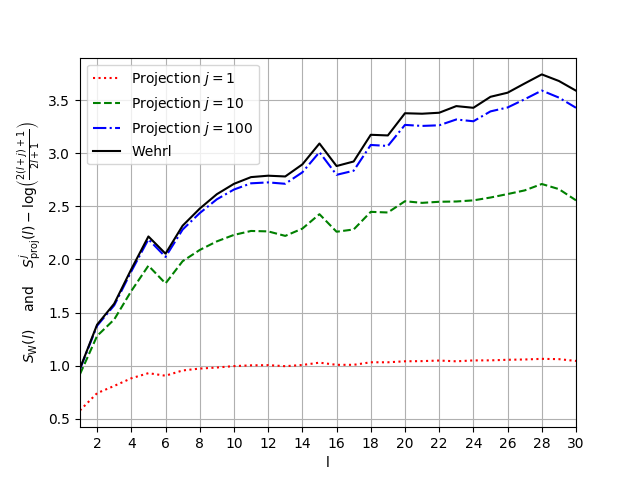}
\caption{Comparison of the Wehrl entropy $S_{\mathrm{W}}(l)$ with the $j=1$-,$10$-,$100$-projection pseudoentropies minus a $j$- and $l$-dependent term, $S_{\mathrm{proj}}^{(j)}(l) + \ln\left(\frac{2l+1}{2(l+j)+1}\right)$, for the NILC 2015 map on the range $[1,30]$. For $j \rightarrow \infty$ the latter converges to the former, but not uniformly.}
\label{ProjWehrl}
\end{center}
\end{figure}
In the limit $j \rightarrow \infty$ this inequality becomes an equality, see Fig.~\ref{ProjWehrl} for the converge of $S_{\text{proj}}^{(j)}$ to $S_{\mathrm{W}}$ for Needlet Internal Linear Combination (NILC) \textit{Planck} data and \cite{Lieb91} for an alternative proof. The projector $P_{l+j}: [l] \otimes [j] \rightarrow [l+j]$ can be expressed in terms of Clebsch-Gordan coefficients:
\begin{equation}
P_{l+j} \big( |l,m\rangle \otimes |j,M\rangle\big) = \sqrt{\frac{\binom{2l}{l+m}\binom{2j}{j+M}}{\binom{2(l+j)}{l+j+m+M}}} \ .
\end{equation}
For large $j$ the projection method provides a good way to compute the Wehrl entropy with high precision. For small $j$ we get an entropy measure with all the nice properties of Wehrl entropy, but a pretty large computational advantage. We will now focus on the case where $\rho$ is a pure state, i.e., $\rho = |\psi_l\rangle\!\langle\psi_l|$ with $|\psi_l\rangle$ computed from the $a_{lm}$ of the CMB data with  $l$ fixed. For a pure state $\rho$ the matrix \eqref{projmatrix} can be rewritten as the Gram matrix of a set of vectors $\vec V_M \in [l+j]$ that are labeled by a basis of $[j]$:
\begin{align}
P_{l+j}\left(|\psi_l\rangle\!\langle\psi_l| \otimes \mathbf 1\right) P_{l+j} & = \sum_{M=-j}^j \vec V_M \vec V_M{}^\dagger \\
\text{with} \quad  \vec V_M &= P_{l+j}\big(|\psi_l\rangle\otimes|j,M\rangle\big) \ .
\end{align}
The dual Gram matrix
\begin{equation}
\begin{split}
\mathrm{Tr}_{[l+j]} \big( \vec V_M \vec V_{M'}{}^\dagger \big) &= \vec V_{M'}{}^\dagger  \cdot  \vec V_M \\
&=  \big(\langle\psi_l|\otimes\langle j,M'|\big) P_{l+j} \big(|\psi_l\rangle\otimes|j,M\rangle\big)
 \end{split}
\end{equation}
has the same non-zero eigenvalues as the original matrix, because for any matrix $C$, $C C^\dagger$ and $C^\dagger C$ have the same non-zero singular values. 
We can therefore also use the dual Gram matrix for the computation of the projection entropy. Appropriately normalized and written in basis-independent notation we have
\begin{align}
\tilde \rho_{\text{proj}}^{(j)} &= \frac{2l+1}{2(l+j)+1} \langle \psi_l| P_{l+j}|\psi_l\rangle  \\ 
S_{\text{proj}}^{(j)}(\rho) &= - \mathrm{Tr} \left(  \tilde \rho_{\text{proj}}^{(j)} \ln \left(\tilde \rho_{\text{proj}}^{(j)}\right) \right) \ ,
\end{align}
where the expectation value is taken in the first tensor slot of $P_{l+j}$.
Unlike $\rho_{\text{proj}}^{(j)}$ the new density matrix $\tilde \rho_{\text{proj}}^{(j)}$ is in general not a faithful representation of the underlying $\rho$ for $j < l$, but the entropy is precisely the same, while its computation involves smaller matrices and is faster. The computational advantage is particularly large for small $j$. The projection entropy computed in this way is an excellent tool for the analysis of the CMB and other spherically distributed data. 

Expanding the unit operator on $[j]$ in Eq.~\eqref{mixproj} in terms of basis states, it can be seen that the map $\rho \rightarrow \rho_{\text{proj}}^{(j)}$ is in fact a trace preserving completely positive map (quantum channel) $[l] \rightarrow [l+j]$ in Kraus form:
\begin{align}
\rho_{\text{proj}}^{(j)} &= \sum_M A_M \rho A_M{}^\dagger \ , \quad \sum A_M{}^\dagger A_M = 1 \\
 \quad A_M &= \sqrt{\frac{2l+1}{2(l+j)+1}}P_{l+j} |j,M\rangle \ ,
\end{align}
where the last terms can also be written $P_{l+j} |j,M\rangle = \sum_m |j+l,m+M\rangle\!\langle l, m|$. There is a similar formula for the transformation of the density matrix in the the dual Gram matrix formulation. In view of the $j\rightarrow \infty$ limit, the lower symbol of a density matrix can also be interpreted as resulting from a completely positive map. 

We shall now introduce yet another natural choice of a rotationally invariant quantum channel, leading to what we call ``angular entropy'', which shares the nice properties of the aforementioned entropies with the additional advantage of being even faster to compute. Let $L_1$, $L_2$, $L_3$ be the standard angular momentum generators in the spin-$l$ representation and define a mixed density matrix and entropy via
\begin{align}
\rho_\text{ang} &= \frac{1}{l(l+1)} \sum_{i=1}^3 L_i \rho L_i{}^\dagger \\
 \quad S_\text{ang} &= - \mathrm{Tr}\left( \rho_\text{ang} \ln \left( \rho_\text{ang} \right) \right) \ . \label{angentropy}
\end{align}
The transformation is obviously of Kraus form and therefore completely positive. It is trace-preserving because $C = \sum_i L_i{}^\dagger L_i$ is the quadratic casimir and has value $l(l+1)$ in the spin $l$ representation. The formula for angular entropy can be written in a basis-independent way by replacing 
$\sum L_i \otimes L_i$ by $\tfrac 12 (\Delta C - C\otimes 1 - 1 \otimes C)$, where $\Delta C$  the coproduct of the casimir. In practice the formula is usually rewritten in terms of $\tfrac 1{\sqrt 2} L_\pm$ instead of $L_1$ and $L_2$. Therefore we have included the dagger $\dagger$ in Eq.~\eqref{angentropy}, which is of course not necessary for Hermitian $L_i$.
For a pure state $\rho = |\psi\rangle\!\langle \psi|$, there is also a dual Gram matrix formulation of the angular entropy:
\begin{equation}
\label{ang_alt}
G_{ij} = \langle \psi | C^{-1} L_i{}^\dagger L_j|\psi\rangle \ , \quad S_\text{ang} = - \mathrm{Tr} ( G \ln (G) )\ 
\end{equation}
(see also \cite{PowerEntropyI,PowerEntropyII} for application to CMB data).
In the way we have written this formula, it is now in fact no longer restricted to individual angular momentum (multipole) numbers $l$. It can also be applied to a range or even a selection of $l$: One simply needs to insert an appropriately normalized state
\begin{equation}
\label{psi_range}
|\psi\rangle = \sum_{l \in \text{select}} \sum_m a_{lm} |l,m\rangle \ ,
\end{equation}
leading to the following explicit algorithm: First determine the Hermitian $3\times 3$ dual Gram matrix $G$
\begin{equation}
G = \begin{pmatrix} G_{11} & G_{12} & G_{13}\\ G^*_{12} & G_{22} & G_{23} \\ G^*_{13} & G^*_{23} & G_{33} \end{pmatrix}
\end{equation}
with matrix elements
\begin{equation}
G_{ij} = \sum_{l \in \text{select}} \frac{1}{2l(l+1)}\mathfrak{g}_{ij}(l,\{a_{lm}\}) \, ,
\end{equation}
with 
\begin{align*}
\mathfrak{g}_{11} &= \sum_{m=-l}^l (l+m+1)(l-m) \cdot |a_{lm}|^2 \\
\mathfrak{g}_{12} &= \sum_{m=-l+1}^{l-1}\sqrt{(l^2-m^2)((l+1)^2-m^2)} \cdot a_{l,m-1} a^*_{l,m+1} \\
\mathfrak{g}_{13} &= \sqrt{2}\sum_{m=-l}^{l-1} (m+1)\sqrt{(l+m+1)(l-m)} \cdot a_{lm}a^*_{l,m+1} \\
\mathfrak{g}_{22} &= \sum_{m=-l}^l (l-m+1)(l+m) \cdot |a_{lm}|^2 \\
\mathfrak{g}_{23} &= \sqrt{2}\sum_{m=-l+1}^l (m-1) \sqrt{(l-m+1)(l+m)} \cdot a_{lm}a^*_{l,m-1} \\
\mathfrak{g}_{33} &= 2 \sum_{m=-l}^l m^2 \cdot |a_{lm}|^2 \ ,
\end{align*}
where ``select'' is a chosen selection of multipole angular momentum quantum numbers~$l$. In this paper we typically select a single value at a time, but this can also be a range of values or an even more complex selection. The angular entropy is then computed in terms of the three eigenvalues $\lambda_i \in [0,1)$ of 
 the normalized mixed angular density matrix $\rho_\text{ang} =  G/{\mathrm{Tr} (G)}$, with $\mathrm{Tr}(G) = \sum_{l \in \mathrm{select}} (2l+1) \hat{C}_l$,
\begin{equation}
\label{ang_range}
S_\text{ang} = - \mathrm{Tr}( \rho_\text{ang}  \ln \left(\rho_\text{ang}\right)  )= - \sum \lambda_i \ln (\lambda_i) \ 
\end{equation}
(see \cite{Kopp} for an overview of algorithms for the fast and precise computation of eigenvalues of Hermitian $3 \times 3$ matrices). There exist also two range entropy measures using the Wehrl entropy which were identified in \cite{Fintzen}.

The computation of the angular entropy involves only $3 \times 3$ matrices and their eigenvalues. It is by far the fastest method and numerical experiments with actual and simulated data show that it has similar behavior as the Wehrl entropy. 
From a theoretical point of view there are some similarities between the angular and projection entropy and hence also the Wehrl entropy: As we have mentioned, the projector $P_{l+j}$ is related to the Clebsch-Gordon decomposition of $[l] \otimes [j]$. Likewise, the angular momentum operators can be interpreted as Clebsch-Gordon coefficients of the decomposition of $[l] \otimes [l]$, but while the projector is onto the highest spin component, the angular momentum generators pick out the adjoint spin-$1$ representation. For the angular entropy there are similar conjectures as for the Wehrl and projection entropies, which are still open and under current consideration. There are several further generalized pseudoentropies -- for example one could choose a (convex) function of the casimir in the definition of angular entropy, e.g.
\begin{equation}
\frac{1}{l(l+1)}\sum_{i_1,\ldots,i_k}L_{i_1}\cdots L_{i_k} \rho L_{i_k} \cdots L_{i_1} \ .
\end{equation}
Below, we focus on the projection and angular entropies that we have defined in this section.

\subsection{Probability distribution of the angular pseudoentropy}

In this section, we want to compare the behavior of the angular pseudoentropy for isotropic and Gaussian maps to its behavior for maps that are constructed from multipole vectors which are distributed uniformly on the sphere according to the surface measure. The latter are statistically isotropic but not Gaussian, hence we investigate deviations from Gaussianity without violating statistical isotropy. For later convenience we often use the logarithmic reciprocal distance of the angular pseudoentropy from its theoretical maximum $X:=\ln\left( \frac{1}{\ln(3)-S_{\mathrm{ang}}} \right)$ as the quantity in investigation. Whenever $X$ is plotted as the independent variable, the dependent probability and cumulative densities are meant to be $p_X$ and $F_X$, and not $p_S$ and $F_S$. On the other hand, if we plot $S$ as the independent variable, the dependent densities are $p_S$ and $F_S$. The probability densities $p_S$ and $p_X$ are related via Eq.\,\eqref{analytical_l2_a}. Note that $X(S)$ is a monotonic function and hence minimal/maximal $S$ corresponds to minimal/maximal $X$. Since we do not take into account the other entropies in this section, we drop the subscript 'ang' in the text.

In the future one should also consider small deviations from isotropy and Gaussianity and investigate the behavior of the entropy distribution in dependence on the small deviation parameters. In this work we leave it at the most simple deviation from Gaussianity in the form of uniform MPVs but consider maps which are constructed from partly Gaussian $a_{lm}$ and partly uniform MPVs as well. Our main aim is to show, that there is a distinction between Gaussian and non-Gaussian maps in the entropy statistics. It should be noted that Gaussianity and statistical independence of the $a_{lm}$ for given $l$ are quite closely related and that a major cause of a deviation in the entropy could result from statistical independence of the $a_{lm}$.

\subsubsection{Semi-analytical distribution for uniform MPVs at $l=2$}

\begin{figure}
\includegraphics[width = 0.49\textwidth]{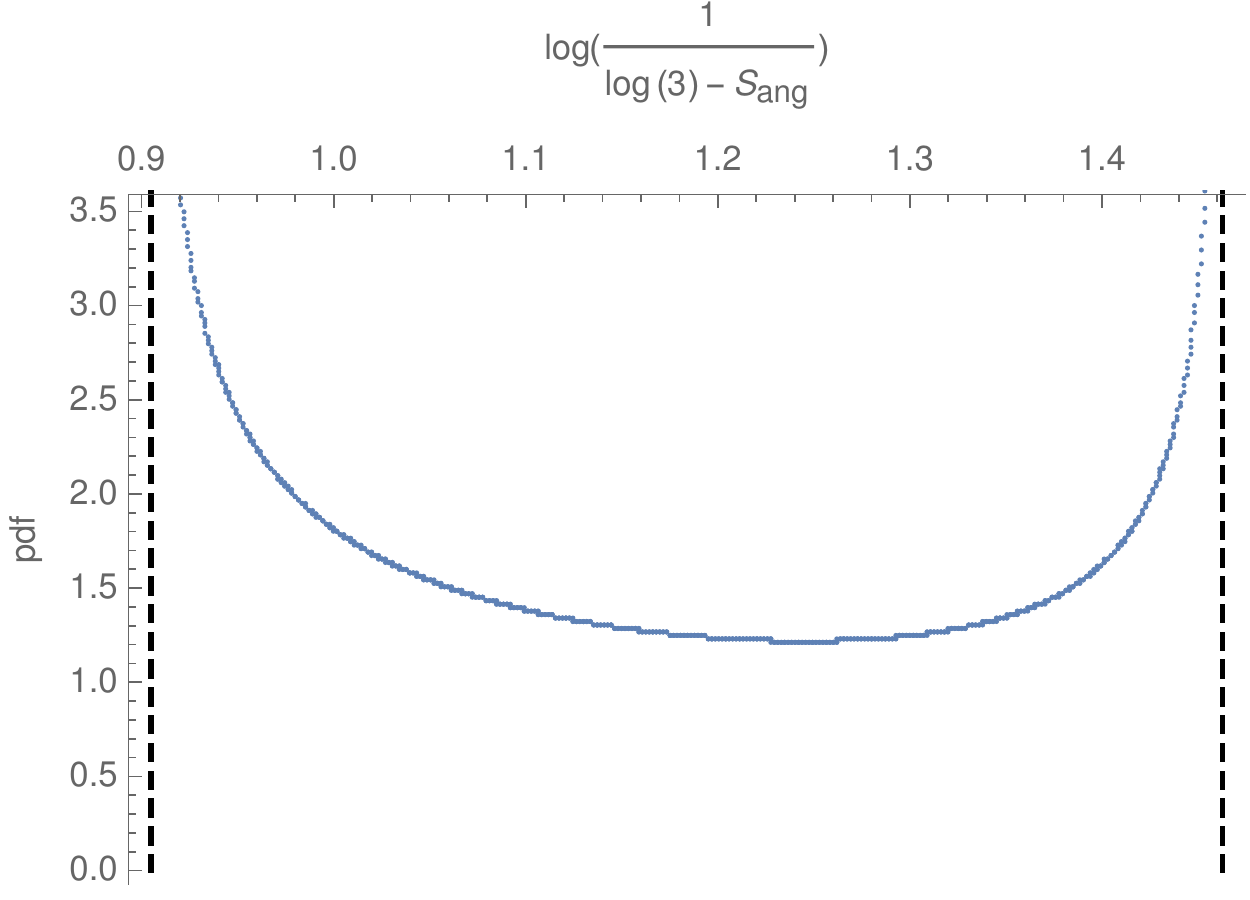}
\caption{Probability distribution for uniformly distributed multipole Vectors of the logarithmic reciprocal of the entropy at $l=2$; calculated by solving $S(\epsilon) = s$ for $\epsilon$ numerically with $S(\epsilon)$ given in \eqref{Sang(eps)} and using the analytical formulas \eqref{analytical_l2_a} and \eqref{analytical_l2_b}. The dotted vertical lines display the asymptotics.}
\label{l2_theo}
\end{figure}

For $l=2$, we have the analytical formula \eqref{Sang(eps)} which expresses the angular pseudoentropy as a function of the squared chordal distance $\epsilon$ between multipole vectors. This can be used to obtain an expression for the probability distribution of $X=\ln\left( \frac{1}{\ln(3)-S} \right)$ if the probability distribution of $\epsilon$ is known. If we consider uniformly distributed multipole vectors on the sphere, which yield an isotropic but non-Gaussian map, the ($l=2$)-case is particularly simple. One can fix the first MPV to be $(0,0,1)^T$ and the second to be an arbitrary vector with length $1/2$ and $z\geq 0$. Then for the angle $\Theta$ between both we have $p^{\mathrm{uni}}_{\Theta}(\theta) = \sin(\theta)$ and since $\epsilon(\Theta) = \sin^2(\Theta/2)$, the probability distribution for $\epsilon$ is $p^{(\mathrm{uni},2)}_{\epsilon}(\epsilon) \equiv 2$. This induces the following probability distributions for $S$ and $X$:
\begin{align}
\label{analytical_l2_a}
p^{(\mathrm{uni},2)}_X(x) &= e^{-x} p^{(\mathrm{uni},2)}_S\left(\ln(3)-e^{-x}\right) \\
p^{(\mathrm{uni},2)}_S(s) &= \frac{2}{\frac{\mathrm{d}S}{\mathrm{d}\epsilon}|_{\epsilon(s)}}.
\label{analytical_l2_b}
\end{align}
Unfortunately, $S(\epsilon) = s$ is a transcendental equation and therefore has to be solved for $\epsilon(s)$ numerically. Fig.~\ref{l2_theo} shows $p^{(\mathrm{uni},2)}_X(x)$. Large and small values of the pseudo angular entropy are preferred in this case, because the slope of $S(\epsilon)$ approaches zero in these regimes. Since $S(\epsilon)$ is compactly supported, so is $p_X^{(\mathrm{uni},2)}$. As is shown below, the case $l=2$ is special among all multipoles.

\subsubsection{Numerical distributions at $l>2$}

\begin{figure}
\begin{subfigure}[c]{0.49\textwidth}
\includegraphics[width = \textwidth]{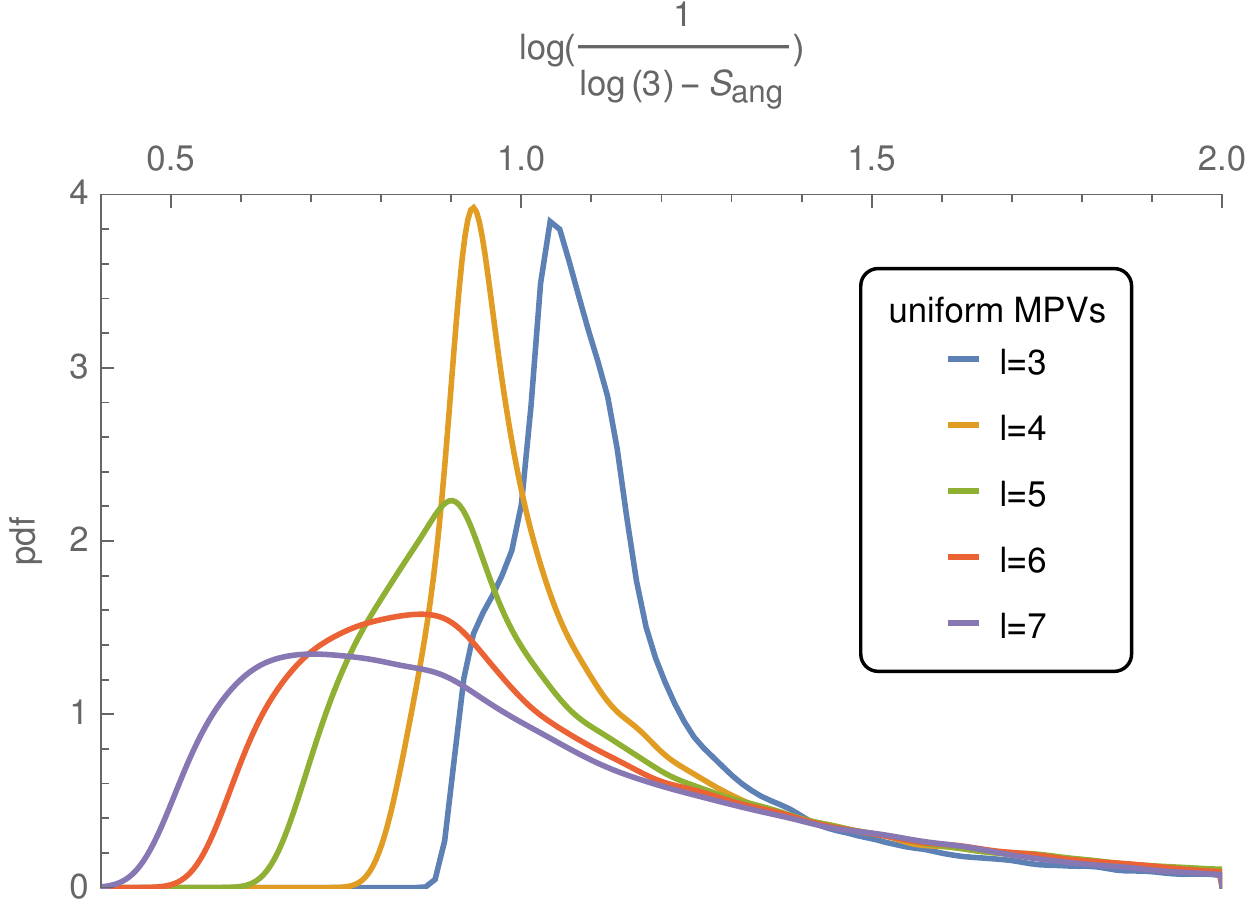}
\end{subfigure}
\begin{subfigure}[c]{0.49\textwidth}
\includegraphics[width = \textwidth]{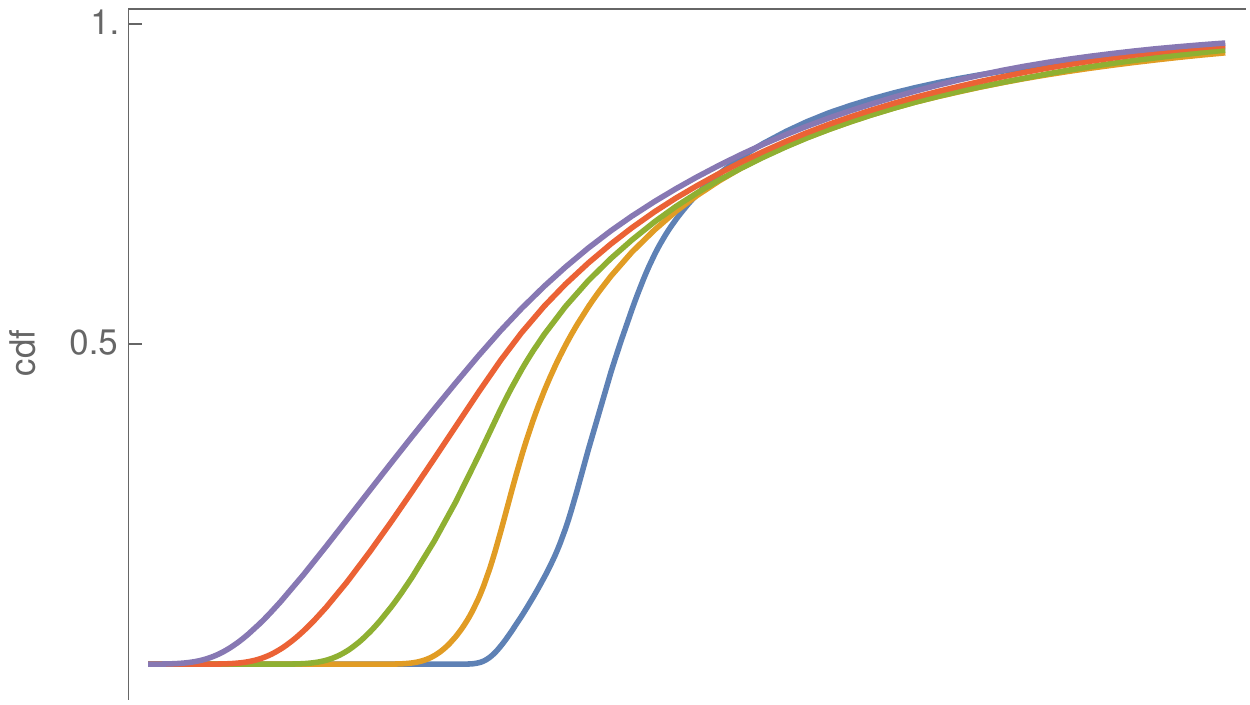}
\end{subfigure}
\caption{Probability (top) and cumulative (bottom) distribution of the logarithmic reciprocal of the angular pseudoentropy for uniformly distributed multipoles vectors at multipoles $l=2,3,4,5,6,7$; calculated with $10^5$ random ensembles and smoothed.}
\label{mpv_pdf_cdf}
\end{figure}

\begin{figure}
\begin{subfigure}[c]{0.49\textwidth}
\includegraphics[width = \textwidth]{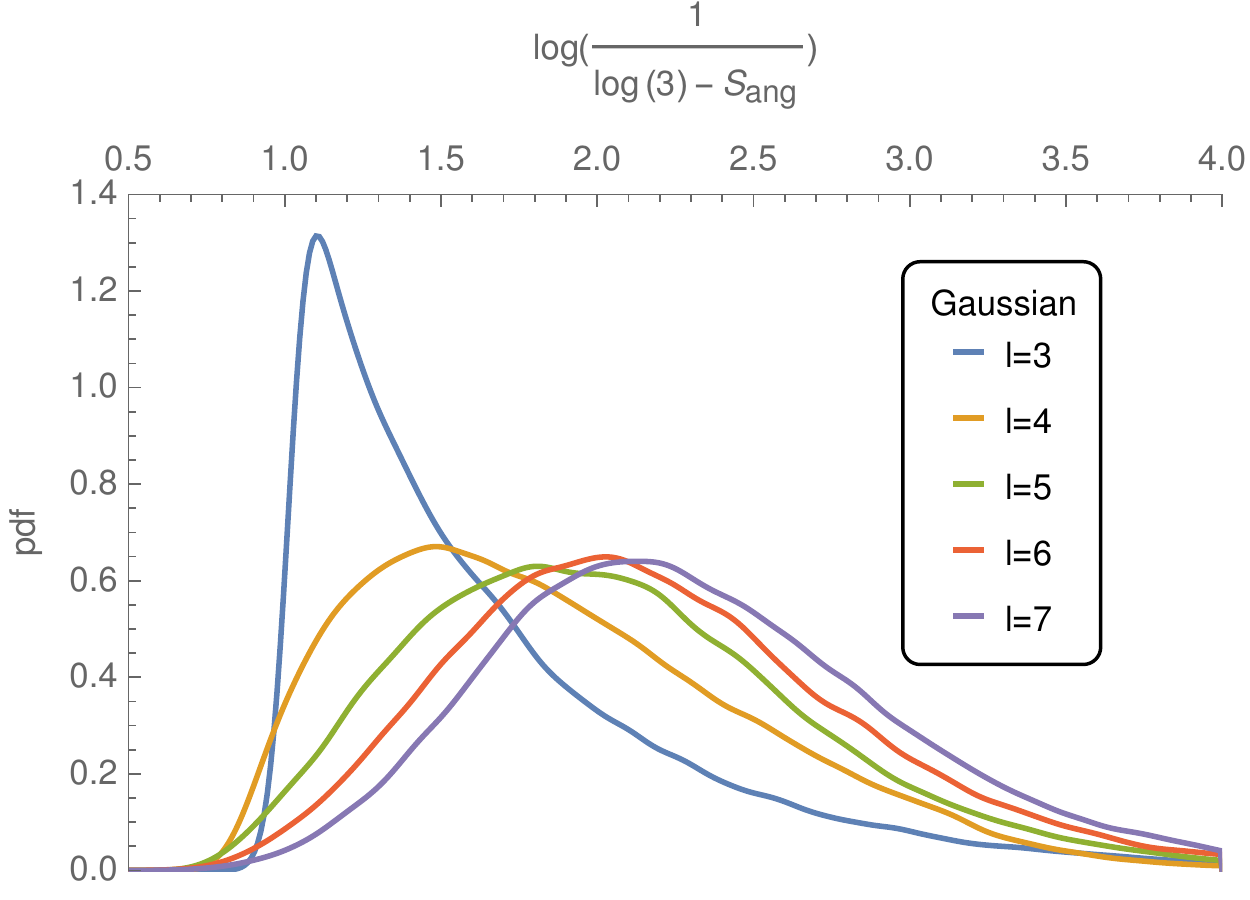}
\end{subfigure}
\begin{subfigure}[c]{0.49\textwidth}
\includegraphics[width = \textwidth]{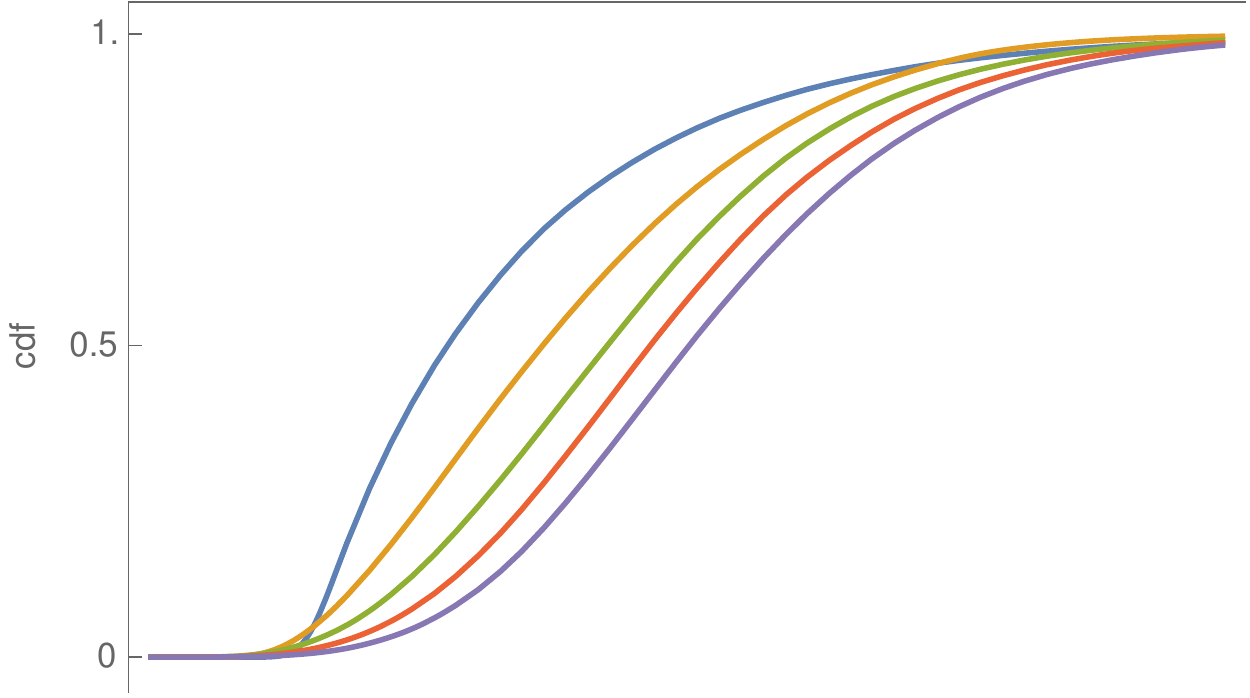}
\end{subfigure}
\caption{Probability (top) and cumulative (bottom) distribution of the logarithmic reciprocal of the angular pseudoentropy for Gaussian and isotropic $a_{lm}$ at multipoles $l=2,3,4,5,6,7$; calculated with $10^5$ random ensembles and smoothed.}
\label{pdf_cdf_ang}
\end{figure}

\begin{figure}[h]
\begin{subfigure}[c]{\columnwidth}
\includegraphics[width = \columnwidth]{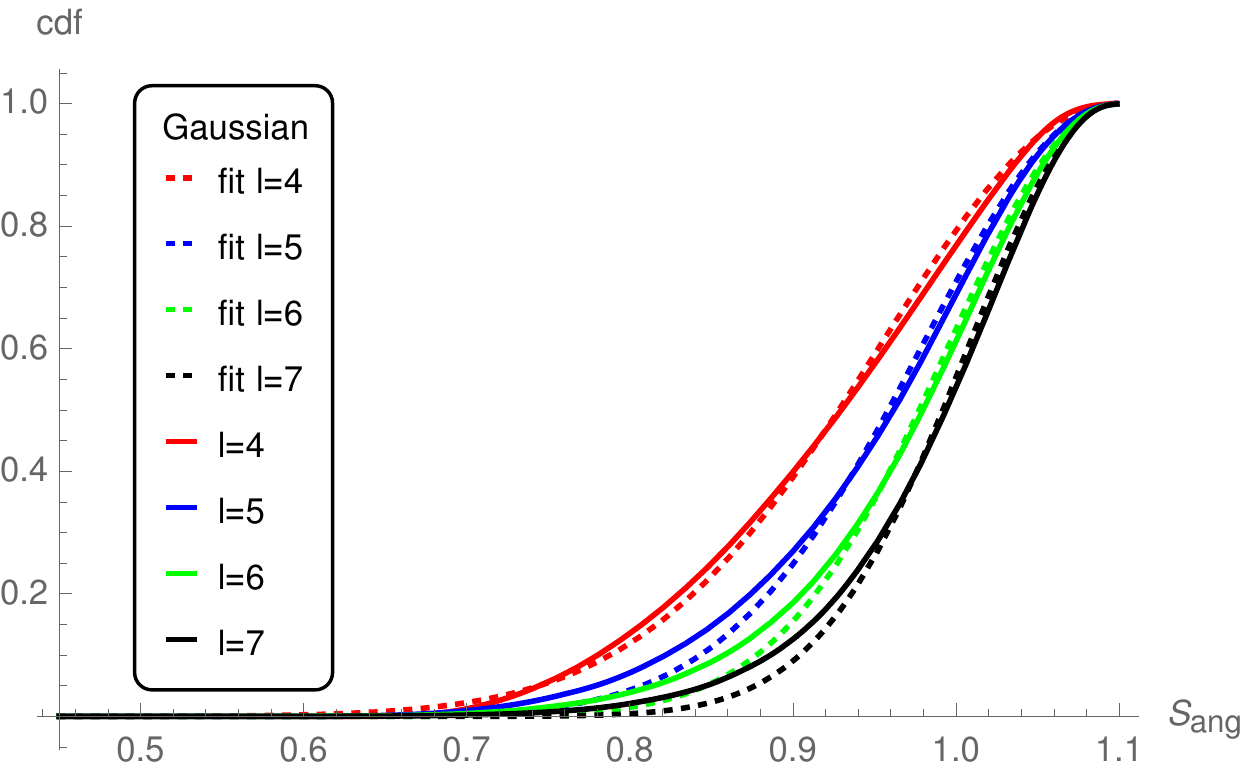}
\end{subfigure}
\begin{subfigure}[c]{\columnwidth}
\includegraphics[width = \columnwidth]{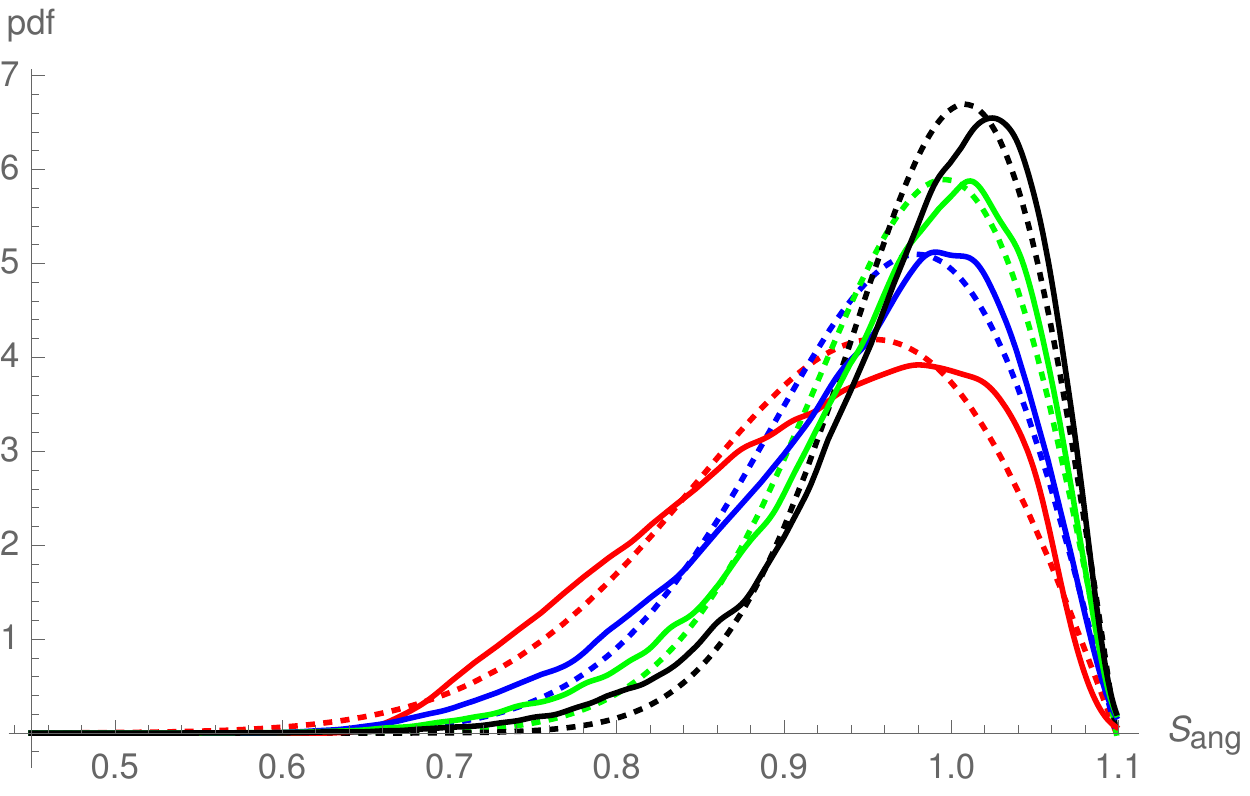}
\end{subfigure}
\caption{Cumulative (top) and probability (bottom) distribution of the angular pseudoentropy for isotropic, Gaussian $a_{lm}$ at $l=4,5,6,7$; calculated with $10^5$ random ensembles and smoothed. The cumulative distribution was fitted with $f(x;a) = e^{-a(x-\log(3))^2}$ and is shown together with the fit functions. The probability distribution is shown together with the derivatives of the fit functions $f'(x;a)$.}
\label{prob_fit}
\end{figure}

\begin{figure}
\begin{subfigure}[c]{\columnwidth}
\includegraphics[width = \textwidth]{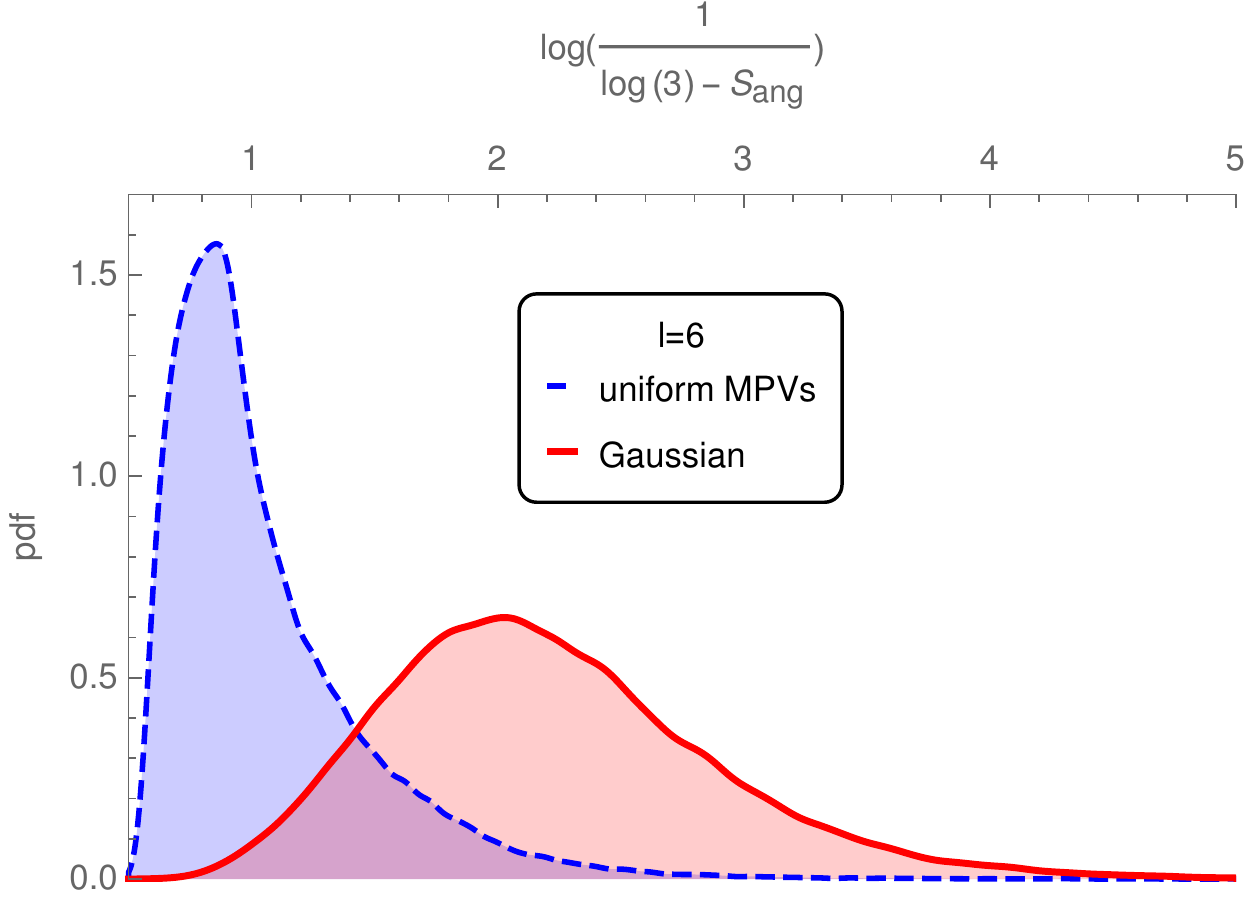}
\end{subfigure}
\begin{subfigure}[c]{\columnwidth}
\includegraphics[width = \textwidth]{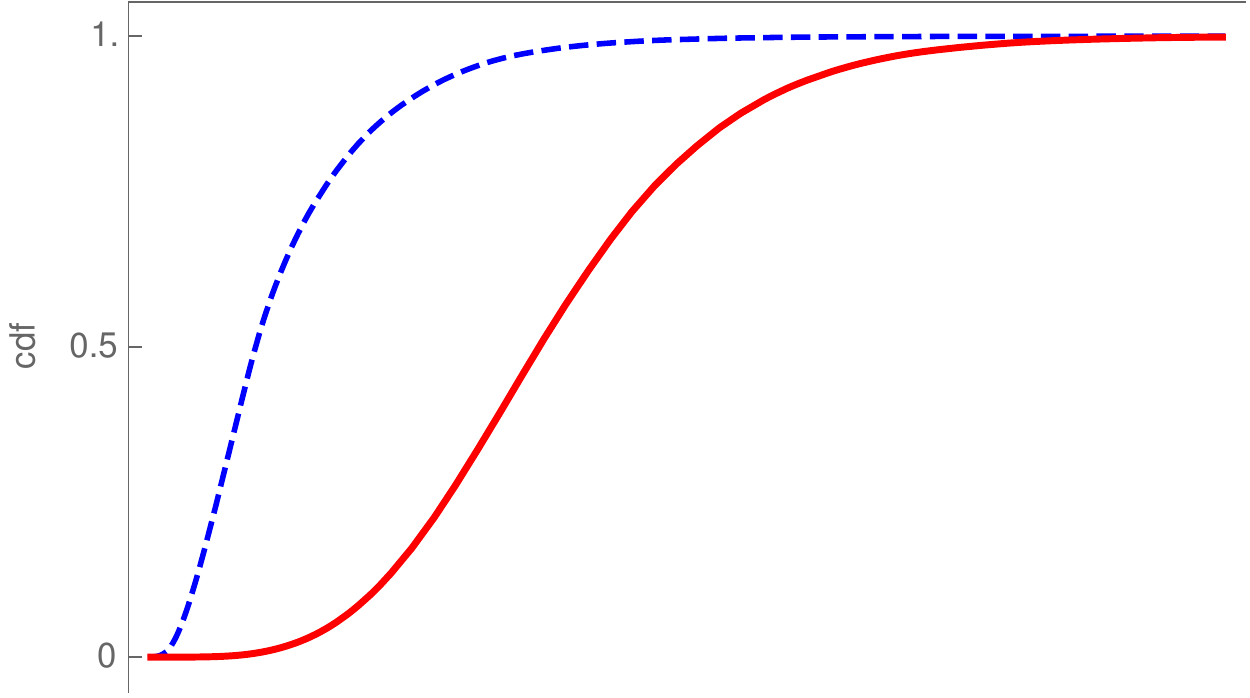}
\end{subfigure}
\caption{Comparison between isotropic Gaussian $a_{lm}$ and uniformly distributed multipole vectors of probability (top) and cumulative (bottom) distribution of the logarithmic reciprocal of the angular pseudoentropy at $l=6$; calculated with $10^5$ random ensembles and smoothed.}
\label{mpv_vs_gaussian}
\end{figure}

For $l>2$ the analytical result for $S$ is a complicated expression and therefore we resort to Monte Carlo simulations. We computed probability and cumulative distributions for $l=2$ (Fig.~\ref{l2Gauss} in Appendix \ref{app_plots}) as well as $l=3$--$7$ with $10^5$ random ensembles (Figs.~\ref{mpv_pdf_cdf},\ref{pdf_cdf_ang},\ref{prob_fit},\ref{mpv_vs_gaussian}) and for $l=20,40,60,80,100$ with only $10^2$ random ensembles (see Figs.~\ref{prob_highl},\ref{l100} in Appendix \ref{app_plots}). The angular pseudoentropy is capable of distinguishing clearly between isotropic Gaussian maps and isotropic non-Gaussian maps connected with uniformly distributed MPVs, especially at high $l$, but not at $l=2,3$.

Figure \ref{mpv_pdf_cdf} shows the distributions for uniform MPVs at large angular scales. For increasing $l$, the distribution gradually moves to smaller entropy values. This behavior carries on to larger multipole numbers $l\leq 100$ (see Fig.~\ref{prob_highl}). The large entropy behavior shows up to be universal on the range $l \in [3,7]$. Due to the low number of ensembles, we cannot confirm this property for larger $l$, but we observe that the right tail does not stretch further out and hence is bounded from above by the right tail at lower multipoles. This means that from $X = 2.5$ on the probability distribution for uniform MPVs is effectively zero.

The distributions for isotropic and Gaussian maps at large angular scales (without $l=2$) are shown in Fig.~\ref{pdf_cdf_ang} and the distribution at $l=2$ in Fig.~\ref{l2Gauss} in Appendix \ref{app_plots}. For $l=2$ the distribution peaks at $X \approx 1.5$, decreases towards smaller entropy values and becomes zero at $X \approx 0.9$. The reason for this behavior is that $S$ is in general tightly bounded at the dipole and that MPVs from Gaussian and isotropic maps tend to repel each other. If only two MPVs are present, the most likely configuration is that of orthogonal MPVs, which results in a maximal $S$. For larger $l$, i.e., a higher number of MPVs, the number of configurations that admit a maximal distance increases and hence the distribution is smoothed. $l=3$ is a transition multipole between the smooth and stretched higher multipoles and the sharp and restricted $l=2$. From $l=3$ on, the distribution moves to larger entropy values, which is confirmed at higher multipoles in Fig.~\ref{prob_highl}. The general shape and the width of the probability distribution is approximately conserved when changing $l$ (except for $l=2,3$), only the expectation value is shifted. Hence also confidence levels in $X$ are approximately constant (see also Fig.~\ref{Fit_plot}).

One can try to fit the cumulative distributions of $S$ for isotropic, Gaussian data with a simple function. The comparison between an $e^{-a(x-\ln(3))^2}$-fit and the cumulative distribution as well as between the derivative of the fit function and the probability distribution (see Fig.~\ref{prob_fit}) shows good agreement at the right tail and moderate agreement at the left tail. We conclude that a Gaussian form of the cumulative distribution provides a good first guess also for the probability distribution but should be refined to arrive at a better agreement at the left tail. It should be noted that no analytical result for the left tail is known. Already the calculation of the general lower bound of the entropies is a difficult mathematical problem whose solution for the Wehrl entropy took several decades. Nevertheless, since the whole distribution moves in shape to the right, also the left tail moves to the right when increasing $l$.

Comparing the distributions, one observes that for increasing $l$ the entropy decreases for uniform MPVs and increases for isotropic, Gaussian maps. While the overlap at $l=6$ is already small but could still have an effect (see Fig.~\ref{mpv_vs_gaussian}), the distributions are clearly distinguished at $l=100$ (see Fig.~\ref{l100}). The distinction between Gaussian and non-Gaussian maps improves for increasing multipole number.

For a quantitative estimate of the behavior of the angular pseudoentropy when only small deviations from Gaussianity are considered, one can investigate the probability distribution using maps that are constructed partly from uniform MPVs and partly from MPVs that are extracted from an isotropic and Gaussian map (see Fig.~\ref{GaussAndUni} in Appendix \ref{app_plots} for $l=6$). It is shown that already a small deviation from Gaussianity in the form of a Gaussian map with one single MPV replaced by a uniformly distributed MPV yields a sizable deviation in the probability distribution and that the distribution converges to the distribution for uniform MPVs rapidly when the number of uniform MPVs is increased. Hence, the entropy measure is highly sensitive to non-Gaussianity. A different but numerically more complicated approach would be to consider the convex combination of the isotropic, Gaussian joint probability distribution of spherical harmonic coefficients and an non-Gaussian distribution. This would have the advantage that the convex deviation parameter could be arbitrarily tuned but it would have the disadvantage of arbitrariness in the choice of the added non-Gaussian contribution. We postpone such an investigation to later works.
 
It should be noted that an equivalent expression to the angular entropy has already been introduced under the name of power entropy in \cite{PowerEntropyI} but without reference to the Wehrl entropy and completely positive maps. Furthermore, that work made the wrong assumption that the maximal entropy value $\ln(3)$ would be obtained for isotropic maps. The method was applied to \textit{Planck} and \textit{WMAP} in \cite{PowerEntropyII} but with the main focus on the correlation of multipoles with the quadrupole. There, no large-scale anomalies were observed, but correlations with the quadrupole were found on a wider range of scales.

\section{Application to CMB data}
\label{Results}

We use \textit{Planck} 2015 second release data, in particular, the four cleaned full sky maps COMMANDER, NILC, Spectral Estimation via Expectation Maximisation (SEVEM), and Spectral Matching Independent Component Analysis
 (SMICA), together with the \textit{WMAP} 7-year ILC cleaned full sky map. The names stand for different cleaning algorithms applied to the original data. COMMANDER uses astrophysical models in order to fill in masked regions that contain foreground contamination, NILC stands for "Needlet Internal Linear Combination" and represents a refinement of the ILC algorithm using needlets in harmonic space, SEVEM uses template fitting and SMICA fills masked regions by a Metropolis Monte Carlo random process. Later in this section we also compare the 2015 results we obtain with results obtained from recently published 2018 \textit{Planck} data. We process the data using the Healpy\cite{Healpix} and Numpy packages for Python 2.7. In order to compute confidence levels, a number of ensembles of Gaussian and isotropic random $a_{lm}$ are treated as input
data for the various entropies. Depending on the entropy the number of ensembles ranges from $30$ to $10^4$. 

\subsection{Comparison of pseudoentropies}

\begin{figure}
\begin{center}
\includegraphics[width = \columnwidth]{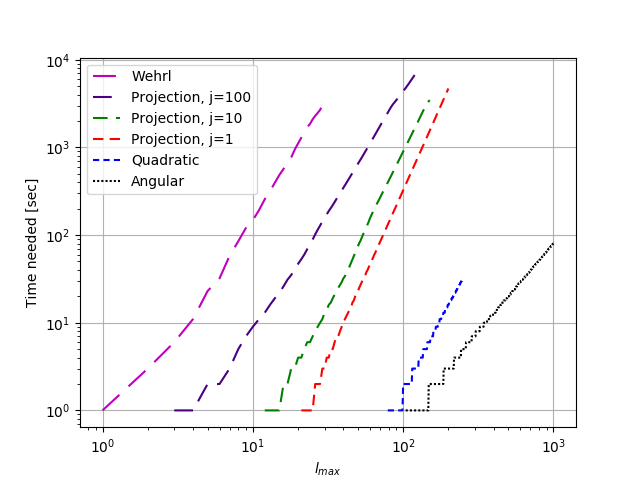}
\caption{Comparison of running times on a standard home computer for one set of $a_{lm}$ when computing the pseudoentropies from $l=1$ up to $l=l_{max}$. For system resources see Appendix \ref{system}.}
\label{Time}
\end{center}
\end{figure}

The considered pseudoentropies differ in computational expense (see Fig.~\ref{Time}). Computing the angular entropy up to $l=1000$ takes about $90$ seconds per run, while the quadratic entropy is slightly more slow. The quadratic entropy is also fast to compute, but it should be used with care since, due to its $-x^2$ instead of the usual $x \log(x)$ behavior it lacks some of the usual entropy properties. Because the projection entropy converges to the Wehrl entropy for $l \rightarrow \infty$ up to a term which does not depend on the data [see Eq.~\eqref{proj_wehrl_conv} in Sec.\ref{math}] its running time converges as well. Clearly, the Wehrl entropy is the quantity that needs the largest computation time, namely about $3000$ seconds up to $l_{max} = 30$ (for system resources, see Appendix \ref{system}).

\begin{figure*}
\begin{subfigure}[c]{0.49\textwidth}
\includegraphics[width = \textwidth]{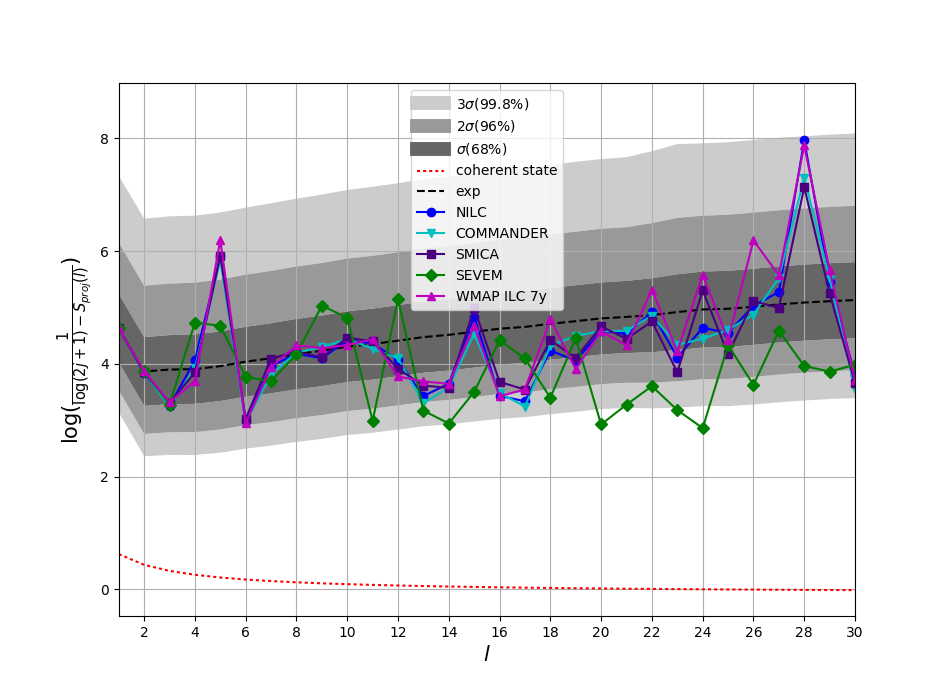}
\subcaption{($j=1$)-projection entropy, 10000 ensembles of random $a_{lm}$.}
\end{subfigure}
\begin{subfigure}[c]{0.49\textwidth}
\includegraphics[width = \textwidth]{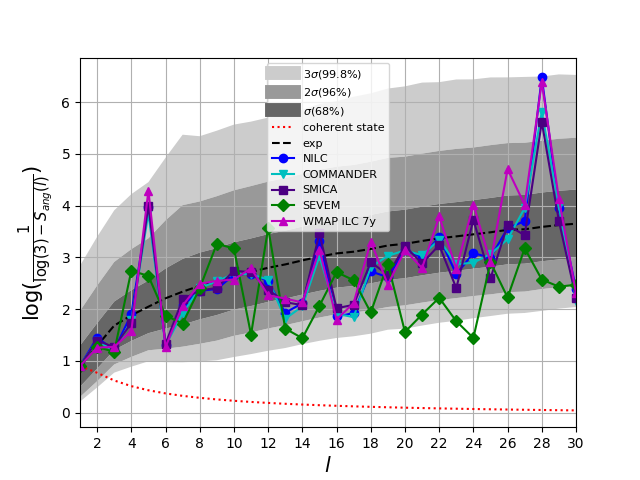}
\subcaption{Angular entropy, 10000 ensembles of random $a_{lm}$.}
\label{ang30}
\end{subfigure}
\begin{subfigure}[c]{0.49\textwidth}
\includegraphics[width = \textwidth]{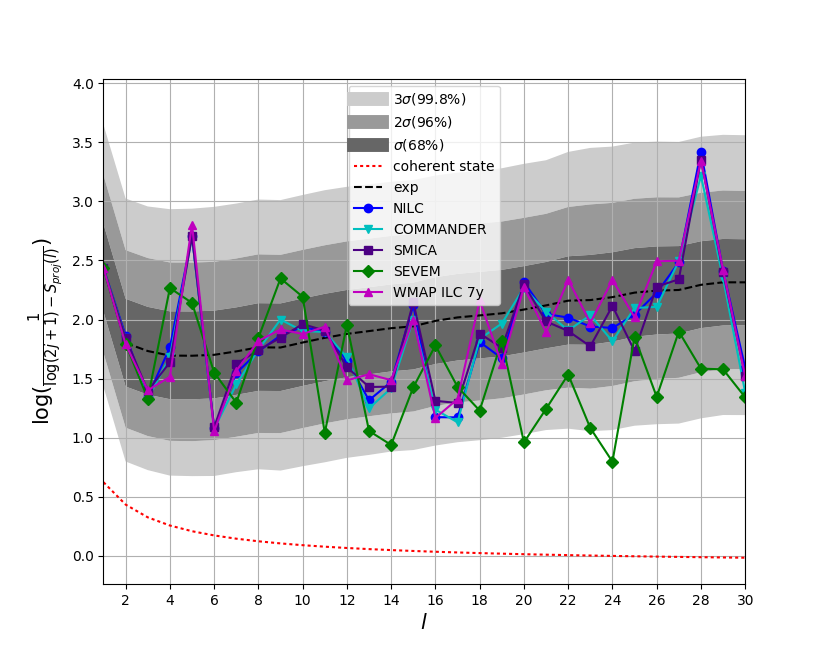}
\subcaption{($j=10$)-projection entropy, 1000 ensembles of random $a_{lm}$.}
\end{subfigure}
\begin{subfigure}[c]{0.49\textwidth}
\includegraphics[width = \textwidth]{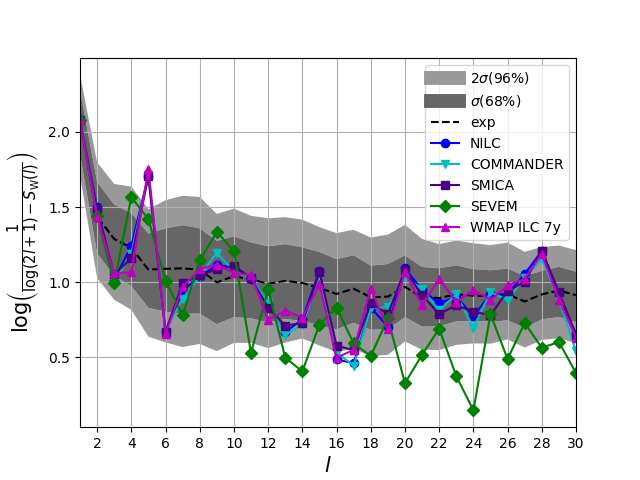}
\subcaption{Wehrl entropy, 30 ensembles of random $a_{lm}$.}
\label{wehrl_30}
\end{subfigure}
\begin{subfigure}[c]{0.49\textwidth}
\includegraphics[width = \textwidth]{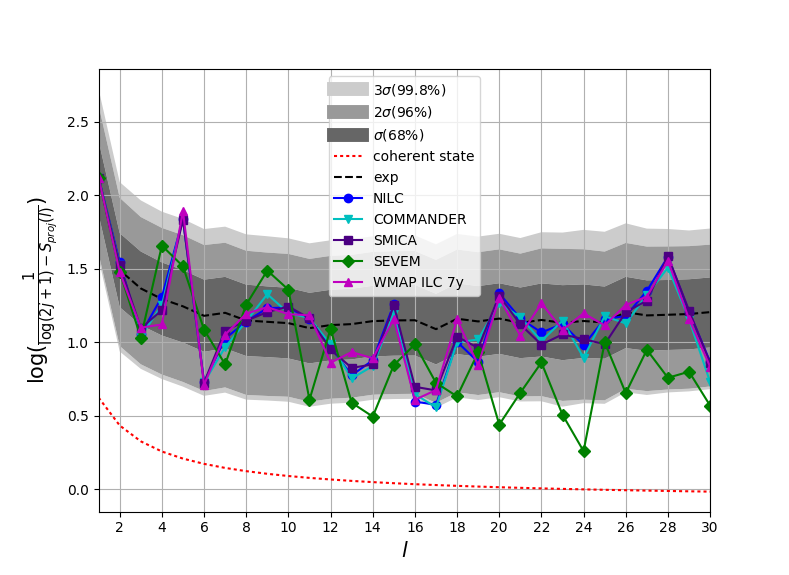}
\subcaption{($j=100$)-projection entropy, 100 ensembles of random $a_{lm}$.}
\end{subfigure}
\begin{subfigure}[c]{0.49\textwidth}
\includegraphics[width = \textwidth]{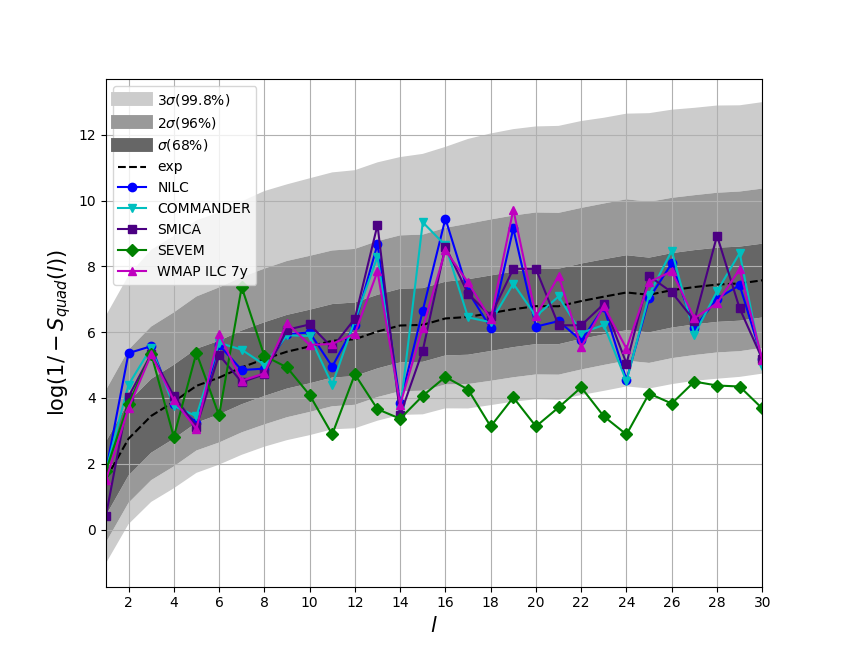}
\subcaption{Quadratic entropy, 1000 ensembles of random $a_{lm}$.}
\label{quad_30}
\end{subfigure}
\caption{Comparison of pseudoentropies from $l=1$ to $l=30$. The sigma boundaries were determined by a certain number of random maps and have been smoothed with a Gaussian filter in all plots. Note, that concerning the angular entropy the smoothing broadens the confidence levels for $l=2,4$ and straightens them for $l=3$. For higher multipole numbers the smoothing does not add or remove any features.}
\label{Ent_comp}
\end{figure*}

Figure \ref{Ent_comp} shows that all measures except the quadratic pseudo-entropy exhibit very similar features in the data analysis, which has also been noticed in \cite{Minkov}. In particular, we observe unusually large values at $l=5$ and $28$ and conspicuously small values at $l = 6, 16, 17$ and $30$. On the other hand, the quadratic pseudoentropy singles out other unlikely multipoles, e.g. $l = 14$. This shows again that this measure should be used with care and that the other measures suit our purposes better. It is interesting to see that the most unusual multipoles $l = 5$ and $l = 28$ have an entropy that is far above the expectation value. For non-Gaussian or non-isotropic maps
one would in general expect the entropy to be lower than the expectation, as will be shown later. 
In every plot the \textit{Planck} SEVEM map clearly deviates from the other maps from $l=10$ on, showing values of each entropy which are too small, hence indicating a preferred direction in SEVEM. On the other hand, from the comparison of \textit{WMAP} to the \textit{Planck} maps it becomes clear that \textit{WMAP} has already been fairly accurate on large angular scales, since in all of the entropies the \textit{WMAP} line sticks closely to the \textit{Planck} lines. An analysis of unusual multipoles and the differences of the various maps will be given in Sec.~\ref{results_ang}.

The Gaussian expectation values of the angular and quadratic entropies -- in both cases we plot the logarithm of the reciprocal distance to the theoretical maximum -- as well as the Wehrl entropy are monotonously increasing functions of $l$, approaching the maximal values for $l \rightarrow \infty$, while the projection entropies -- logarithmic reciprocal distance plotted as well -- decrease in the low-$l$-regime and increase for larger values of $l$. The logarithmic reciprocal plotting turns out to be especially useful because the $\sigma$-regions do not decrease for large multipoles in this measure. In fact, as will be shown later in more detail, the confidence levels are constant from intermediate $l$ on, while the Gaussian expectation for the angular entropy shows a very simple functional dependence on $l$ as well. Furthermore, in contrast to unlogarithmic plotting both the upper and lower confidence levels have approximately the same width allowing for a better identification of unusual multipoles. For comparison, see Fig.~\ref{unlog_ang} in Appendix \ref{app_plots}, which shows the pure angular pseudoentropy.

It should be noted that smoothed confidence levels appear only in the plots. When calculating p-Values in this work, they are calculated directly numerically with the data and no smoothing takes place.

Concluding, the agreement of features in the different pseudoentropies suggests considering only the numerically cheapest entropy apart from the quadratic one. Hence, in the following only the angular pseudoentropy will be considered.

\subsection{Results for angular pseudoentropy with 2015 data}
\label{results_ang}
From fig.~\ref{ang30} we read off that for the angular entropy in the range $1\leq l \leq 30$ five NILC data points lie at $2\sigma$ or even outside of it ($l=5,16,17,28,30$), two of which are even close to $3\sigma$ ($l=5,28$). One could now argue, that it is expected that some data points lie at low confidence levels, but a quick estimation shows that the deviations observed here are still unlikely. The probability for five out of 30 data points to lie outside $2\sigma$ approximately equals the Poisson distribution for five events with a mean rate $\lambda = 1.2 = 30\cdot 0.04$, i.e., 
\begin{equation}
P_{\lambda}(5) = \frac{\lambda^{5}}{5!}e^{-\lambda} \approx 0.6 \%,
\end{equation}
implying that the significance of these unlikely data points is above $2\sigma$.

\begin{figure}
\includegraphics[width = \columnwidth]{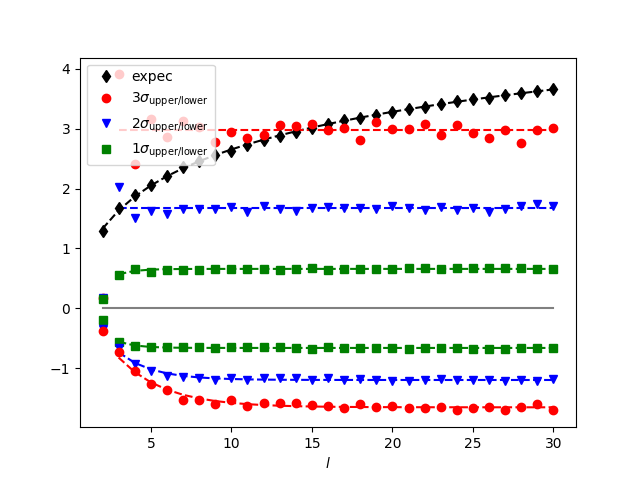}
\caption{Expectation, upper and lower confidence levels of $\log\left(\frac{1}{\log(3)-S_{\mathrm{ang}}(l)}\right)$ with $10000$ ensembles of isotropic and Gaussian random $a_{lm}$ up to $l=100$. The dashed lines represent fits.}
\label{Fit_plot}
\end{figure}

Turning to higher multipole numbers it would be beneficial to find a method of calculating confidence levels even faster. In this regard we observe that in the logarithmic reciprocal depiction the Gaussian expectation value and confidence levels of the angular entropy behave in a simple fashion, namely the expectation can be fitted with $f(x) =a \, \mathrm{log}(bx+c) $ and the confidence levels with $g(x) = a(1-e^{-b(x-1)})$ (see Fig.~\ref{Fit_plot}). In particular, it turns out that the confidence levels are constant from about $l=30$ up to $l=100$. In the following we assume that this holds true for $l>100$. This assumption is justified by continuity of the angular pseudoentropy and the isotropic, Gaussian probability distribution of spherical harmonic coefficients, i.e.,~no sudden jumps should be expected. Tab.~\ref{Fit_params_ang} in Appendix \ref{app_plots} contains all optimal parameters.

While the fit of the expectation value coincides well with the numerical graph on the whole considered range, the lower confidence fits are not suited for $l=2$ and $l=3$ and the upper confidence fits suit the numerical results from $l=3$ ($1\sigma$), $l=4$ ($2\sigma$) and $l=20$ ($3\sigma$). Hence, using the fits in analysis slightly underestimates the most conspicuous multipoles with values above the expectation value in the range $3 \leq l \leq 20$, but the fits allow for a comparison of the entropy to the expectation from $l=1$ to $l=1000$.

\begin{figure}
\begin{subfigure}[c]{0.49\textwidth}
\includegraphics[width = \textwidth]{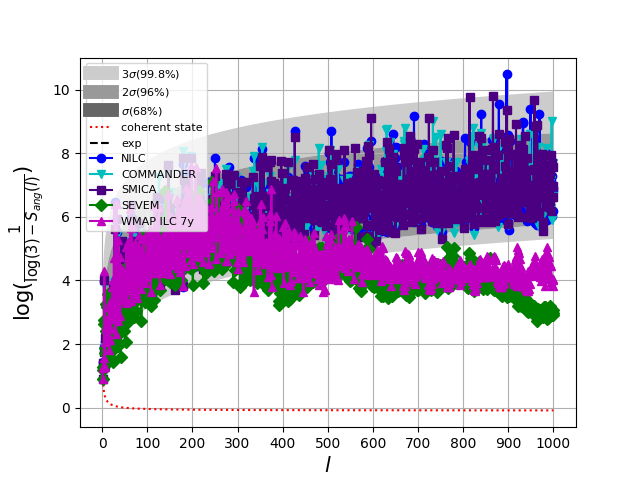}
\end{subfigure}
\begin{subfigure}[c]{\columnwidth}
\includegraphics[width = \textwidth]{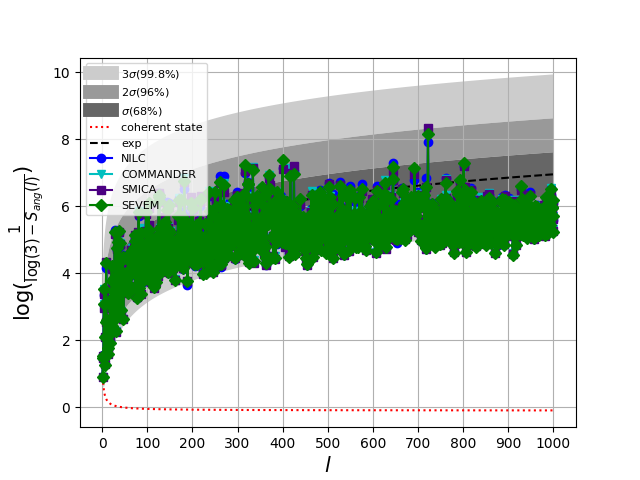}
\end{subfigure}
\caption{Angular pseudoentropy up to $l=1000$ with fitted expectation and confidence levels. Unmasked map (top) and map with SEVEM mask applied and the masked region not filled (bottom).}
\label{ang_1000}
\end{figure}

In Fig.~\ref{ang_1000}, we applied the fits for the angular entropy up to $l=1000$, once for the pure cleaned full sky maps and once masked with the SEVEM mask and without refilling the masked region. In the second case \textit{WMAP} was taken out because of dissimilar NSIDE number of this map and the SEVEM mask. The unmasked shows no obvious deviation of COMMANDER, NILC and SMICA from the expected behavior of a Gaussian map on the whole observed range of multipoles, while single multipoles stick out, as for example NILC at $l=896$, but the data does not exhibit unusual global deviations from the expectation, i.e., deviations on a large range of angular scales. On the other hand \textit{WMAP} and SEVEM clearly fall off from $l=200$ on. While the \textit{WMAP} data is commonly accepted to be inaccurate on very small angular scales, the large drop of SEVEM surprises at first glance. However, the masked plot shows that the deviation of SEVEM from the other \textit{Planck} maps can be explained largely by the strong influence of residual foreground pollution in the SEVEM map. Indeed, the masked \textit{Planck} maps all coincide very well on the whole range, leaving only minor deviations. It can be seen that the pure masking process lowers the entropy for large values of $l$ indicating, as expected, that masking singles out certain directions by removing the galactic plane. This can nicely be seen by taking into account the dashed red line which shows the entropy of a coherent state which represents a map that is confined to a single direction. That masking lowers angular pseudoentropy is not \textit{a priori} clear since we normalize the $a_{lm}$ before computing pseudoentropies and hence there is no lack of absolute power due to masking.

\begin{figure}
\includegraphics[width = \columnwidth]{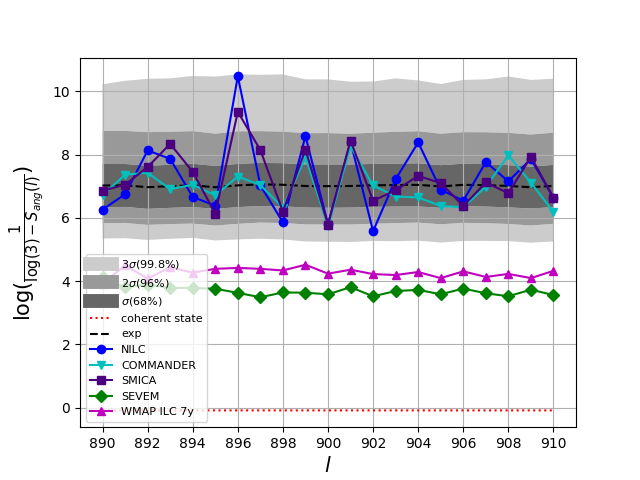}
\caption{Angular pseudoentropy between $l=890$ and $l=910$, confidence levels and expectation calculated with 1000 random ensembles and smoothed with a Gaussian filter.}
\label{ang_high_l}
\end{figure}

\begin{table}
\begin{tabular}{|c||c||c|c|}
\hline
$l$ & pipeline & $p$[\%] & $\bar{p}_{\text{geom}}$[\%] \\
\hline \hline
\multirow{3}{*}{895} & NILC & 18.8 & \\
& COMMANDER & 36.6 & 17.7  \\
& SMICA & 8.1 &  \\
\hline 
\multirow{3}{*}{896} & NILC & 0.1 &  \\
& COMMANDER & 29.5 & 1.2  \\
& SMICA & 0.6 &  \\
\hline 
\multirow{3}{*}{897} & NILC & 46.3 & \\
& COMMANDER & 45.5 & 23.4   \\
& SMICA & 6.1 & \\
\hline 
\multirow{3}{*}{898} & NILC & 2.6 & \\
& COMMANDER & 13.1 & 6.9 \\
& SMICA & 9.7 &  \\
\hline 
\multirow{3}{*}{899} & NILC & 3.1 & \\
& COMMANDER & 10.5 & 5.9 \\
& SMICA & 6.4 &  \\
\hline 
\multirow{3}{*}{900} & NILC & 2.2  & \\
& COMMANDER & 1.1  & 1.7  \\
& SMICA & 2.1 & \\
\hline
\multirow{3}{*}{901} & NILC & 4.1 &  \\
& COMMANDER & 5.4 & 4.3  \\
& SMICA & 3.7 &  \\
\hline 
\multirow{3}{*}{902} & NILC & 0.4 & \\
& COMMANDER & 44.0 & 7.7  \\
& SMICA & 25.9 & \\
\hline 
\multirow{3}{*}{903} & NILC & 36.4  & \\
& COMMANDER & 33.2 & 37.9 \\
& SMICA & 45.1 &  \\
\hline 
\multirow{3}{*}{904} & NILC & 3.0 & \\
& COMMANDER & 31.4 & 13.9 \\
& SMICA & 28.3 &  \\
\hline 
\multirow{3}{*}{905} & NILC & 45.7 & \\
& COMMANDER & 17.3 & 32.3 \\
& SMICA & 42.8 & \\
\hline
\end{tabular}
\caption{P-values of angular pseudoentropy for $895 \leq l \leq 905$, calculated with 1000 ensembles of Gaussian $a_{lm}$, rounded to one decimal place. In the third column the geometric mean of the p-values of the three maps was chosen because of the multiplicative behavior of probabilities.}
\label{pvals_high_l}
\end{table}

Taking a closer look to angular scales around $l=900$ (see Fig.~\ref{ang_high_l} and Table \ref{pvals_high_l}) for the full sky maps reveals that NILC behaves unusually between $l=895$ and $l=905$. We measure unusualness of multipoles with the p-value, using the convention
\begin{align}
p(l) & := \int_{S_{\mathrm{ang}}(l)}^{\mathrm{log}(3)}\! \mathrm{d}s\, p_S(s) \quad \text{if } S_{\mathrm{ang}}(l) > \langle S_{\mathrm{ang}} \rangle \\
p(l) & := \int_{S_{\mathrm{ang}}^{\mathrm{min}}}^{S_{\mathrm{ang}}(l)}\! \mathrm{d}s\, p_S(s) \quad \text{if } S_{\mathrm{ang}}(l) < \langle S_{\mathrm{ang}} \rangle,
\end{align}
where $p_S(s)$ denotes the probability distribution of $S_{\mathrm{ang}}$ for Gaussian and isotropic $a_{lm}$ and $\langle S_{\mathrm{ang}} \rangle$ the respective expectation value. The entropy value at $l=896$ lies outside the $3\sigma$-region with p-value $\lessapprox 0.1$, where the reason for the inequality is the low number (1000) of random ensembles that have been used to calculate this value, which yields a resolution of $0.1$. This means that on average at most one out of 1000 realizations is expected to be larger than the expectation and to be as unusual as or more unusual than the data point. On the other hand one interpretation is that one out of 1000 multipoles is expected to be at least as unusual as the data point. Since $l=896$ is the only NILC data point on $1 \leq l \leq 1000$ that is outside of $3\sigma$, this multipole is still allowed by statistics. Nevertheless, the NILC values of the angular entropy exhibit small p-values at $l=896,898,899,900,901,902,904$. SMICA behaves a bit less extreme than NILC on the considered range and COMMANDER stays inside or close to the $1\sigma$-region with on average large p-values. While for the multipoles $l=898$ to $l=901$ all the pipelines behave similarly, they deviate from each other at the other multipoles between $l=890$ and $l=910$. 

\begin{figure}
\begin{subfigure}[c]{0.49\textwidth}
\includegraphics[width=\columnwidth]{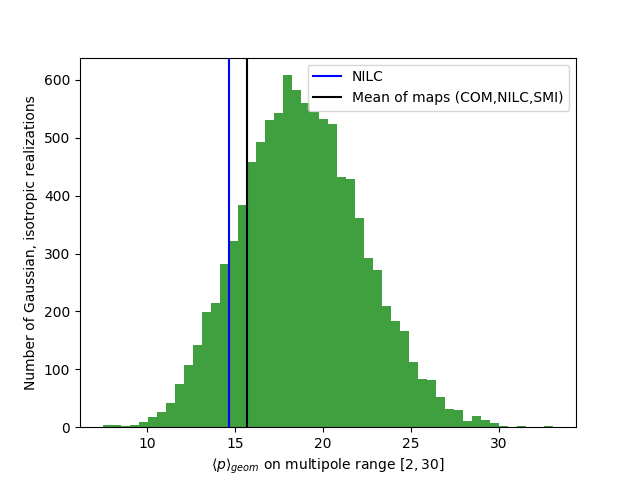}
\subcaption{Range $l\in [2,30]$ for 10000 ensembles of Gaussian and isotropic random maps.}
\label{p_geo_30}
\end{subfigure}
\begin{subfigure}[c]{0.49\textwidth}
\includegraphics[width=\columnwidth]{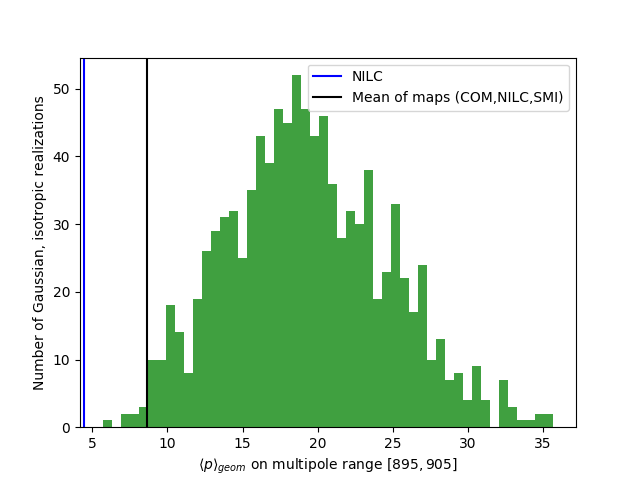}
\subcaption{Range $l\in [895,905]$ for 1000 ensembles of Gaussian and isotropic random maps.}
\label{p_geo_905}
\end{subfigure}
\caption{Probability distribution (unnormalized) of the geometric mean of p-values plotted with 50 bins. The blue and black vertical lines show the NILC p-value and the geometric mean of p-values from COMMANDER, NILC and SMICA for 2015 data.}
\label{p_geo}
\end{figure}

In order to estimate the significance of this multipole range, we calculate the geometric mean over p-values, 
\begin{equation}
\langle p \rangle_{\mathrm{geom}}(\mathrm{map}) = \left(\prod_{l=895}^{905} p^{(\mathrm{map})}(l)\right)^{1/11},
\end{equation}
and compare it to the distribution of p-values for Gaussian and isotropic random maps, see Fig.~\ref{p_geo}. For NILC the geometric mean is $\langle p \rangle_{\mathrm{geom}}(\mathrm{NILC}) = 4.4\%$. From 1000 ensembles of random Gaussian and isotropic maps not a single map attains such a small mean p-value, hence we can give an estimate on the upper bound of the likelihood of the NILC data in the given multipole range assuming Gaussianity and isotropy as a null hypothesis
\begin{equation}
\mathbb{L}(\mathrm{NILC};895 \leq l \leq 905) \lessapprox 0.1\%,
\end{equation}
i.e., an about $3\sigma$-significance. It should be noted that by averaging over a range of $l$-modes one does not take into account correlations of these modes induced by inhomogeneous noise. This effect could cause large upper uncertainties in the likelihood and should be considered seriously in more detailed studies. For judging all three full sky maps together, we use the geometric mean of the p-value over the three maps and proceed with these mean p-values as with NILC, resulting in a mean p-value of $8.6\%$ on $[895,905]$ with a likelihood of
\begin{align}
\begin{split}
&\mathbb{L}(\mathrm{COMMANDER}\cdot\mathrm{NILC}\cdot\mathrm{SMICA}; 895 \leq l \leq 905) \\
 &\approx 0.8\%,
\end{split}
\end{align}
i.e., a more than $2\sigma$-significance. One should keep in mind, that the significance might be lowered when taking into account correlations between the different maps, which are caused by the simple fact, that all of them are derived from the same physical data. Using the geometric mean implicitly assumes that the ingredients are statistically independent. Hence this significance should be seen rather as a first approximation. The intention in taking the geometric mean over different maps is to obtain a p-value which is to some extent independent of the specifics of the different cleaning algorithms.

We conclude that either NILC, and to a lesser extent SMICA, might induce bad characteristics to the data on the mentioned scales or that the COMMANDER algorithm might induce arbitrary isotropy and/or Gaussianity on these scales and therefore distorts the real data. Furthermore, even if one considers all three maps at once by multiplying the p-values and comparing to the expectation, the data is inconsistent with the assumption of isotropy and Gaussianity at a $2\sigma$-level.

It should be noted that this might well be a selection effect due to the particular chosen, non-physically motivated range of scales that was considered. One should ask how likely it is to find a range of multipoles of the given size that which yields an average p-value as low as the considered range. Moreover we do not perform a fully developed and precise statistical analysis, hence the estimated significance might need corrections. The essence here is that the entropy method is capable of highlighting unusual behavior at high multipole numbers.

\begin{table}
\begin{tabular}{|c||c||c|c|}
\hline
$l$ & pipeline & $p$[\%] & $\bar{p}_{\mathrm{geom}}$[\%] (excluding SEVEM) \\
\hline \hline
\multirow{5}{*}{5} & NILC & 1.03 & \\
& COMMANDER & 1.35 &  \\
& SMICA & 0.99 & 0.92 \\
& SEVEM & 16.80 &  \\
& \textit{WMAP} & 0.53 &  \\
\hline 
\multirow{5}{*}{16} & NILC & 1.52 &  \\
& COMMANDER & 2.01 &  \\
& SMICA & 3.58 & 1.89 \\
& SEVEM & 32.10 &  \\
& \textit{WMAP} & 1.16 &  \\
\hline 
\multirow{5}{*}{17} & NILC & 1.66 & \\
& COMMANDER & 1.06 &  \\
& SMICA & 3.82 & 2.35 \\
& SEVEM & 19.50 &  \\
& \textit{WMAP} & 4.56 &  \\
\hline 
\multirow{5}{*}{28} & NILC & 0.07 & \\
& COMMANDER & 0.65 & \\
& SMICA & 1.13 & 0.26 \\
& SEVEM & 4.47 &  \\
& \textit{WMAP} & 0.09 & \\
\hline 
\multirow{5}{*}{30} & NILC & 1.29 & \\
& COMMANDER & 0.32 & \\
& SMICA & 0.45 & 0.63 \\
& SEVEM & 1.93 & \\
& \textit{WMAP} & 0.84 & \\
\hline
\end{tabular}
\caption{P-values of angular pseudoentropy for most conspicuous large angle multipoles, calculated with 10000 ensembles of Gaussian $a_{lm}$, rounded to two decimal places.}
\label{pvals_tab}
\end{table}

Now, we return to the large angular range $l \leq 30$, where we computed p-values with 10000 sets of random $a_{lm}$, see Fig.~\ref{pvals_plot} in Appendix \ref{app_plots} for a plot of the p-values. Tab.~\ref{pvals_tab} shows the six most pronounced large scale multipoles revealing again that unmasked SEVEM does not exhibit the same behavior as the other maps, yielding large p-values at these multipoles while the other maps show small p-values. It turns out that $l=28$ with an average (excluding SEVEM) p-value of about $0.26$\% sticks out most, followed by $l=30$ and $l=5$. Although two of the most conspicuous multipoles display a too large value of the entropy, neither too large nor too small values can directly be identified to be preferred. At these angular scales, the three non-SEVEM \textit{Planck} maps behave quite similarly and the large discrepancy between COMMANDER and NILC is not yet present. Another feature that can be observed is the slight improvement of \textit{Planck} compared to \textit{WMAP}, even for $l < 250$ because on average \textit{WMAP} yields the smallest p-values at these conspicuous multipoles

\begin{figure}
\begin{subfigure}[c]{0.49\textwidth}
\includegraphics[width = \textwidth]{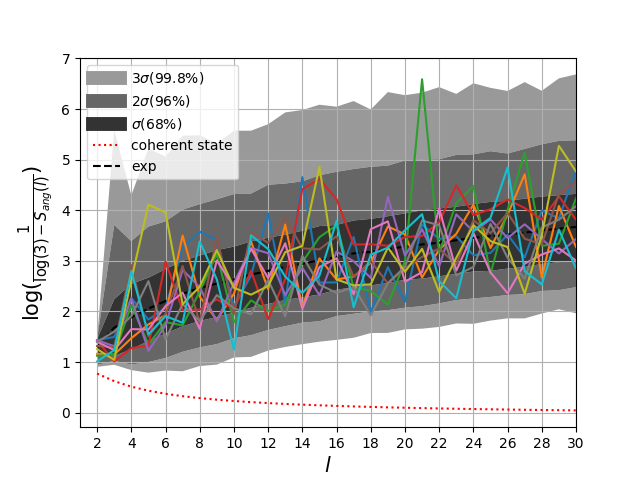}
\subcaption{Unmasked}
\end{subfigure}
\begin{subfigure}[c]{0.49\textwidth}
\includegraphics[width = \textwidth]{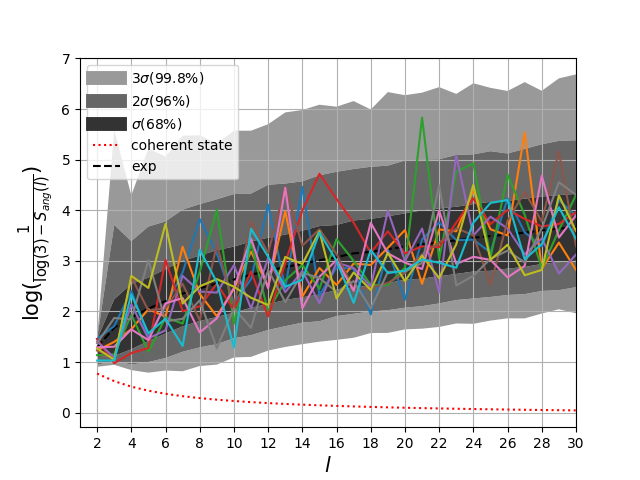}
\subcaption{Masked with \textit{WMAP} intensity mask. The Cosmic Dipole was included before masking and after masking a dipole was removed with Healpy remove\_{}dipole, then the masked region was filled with the original data, see Fig.~\ref{maps} in App.~\ref{app_plots}.}
\end{subfigure}
\caption{Angular pseudoentropy up to $l=30$ from $10$ isotropic and Gaussian random maps with power spectrum (up to $l=200$) of \textit{WMAP} 7-year ILC.}
\label{10_randoms}
\end{figure}

It has been conjectured in \cite{schupp1} that some of the large scale features could be produced by parts of the data processing, namely by a non-linearity in the masking process which mixes the large dipole moment to higher moments when subtracting the dipole. We try to answer this question in a very simplified approach. In Fig.~\ref{10_randoms} we plot the angular entropy for ten isotropic and Gaussian random full sky maps and in a second step we add the non-relativistic contribution of a dipole to the map 
\begin{equation}
T(\vec{e}) \rightarrow T'(\vec{e}) = T(\vec{e}) + A\, \vec{d} \cdot \vec{e},
\end{equation}
where $A = 3364.5 \, \mu\mathrm{K}$ denotes the Cosmic Dipole amplitude and $\vec{d} = (x(l,b),y(l,b),z(l,b))^T$ with $(l,b) = (264.00\deg , 48.24\deg)$ \cite{Planck2015I} denotes its direction in the galactic coordinate system, then we mask the map and remove the dipole afterwards again, using this time the build-in Healpy function \textit{remove\_{}dipole}, which returns a map $\widetilde{T}$ that is the closest -- in the meaning of a least square fit -- map to the original $T$ among those maps obeying $\sum_{p\in\mathcal{P}} \vec{e}_p \widetilde{T}_p = 0$, where $\mathcal{P}$ denotes the set of all unmasked pixels. Finally, we refill the masked region with the original data $T$ in order to receive a full sky map, see Fig.~\ref{maps} in App.~\ref{app_plots} for a depiction of this process by maps in Mollweide view. Since the \textit{WMAP} and \textit{Planck} maps behave similarly on large angular scales and working with \textit{WMAP} is computationally cheaper than working with \textit{Planck} maps -- \textit{WMAP} has $\mathrm{NSide} = 512$ and \textit{Planck} $\mathrm{NSide} = 2048$ -- we use the \textit{WMAP} intensity mask for masking as well as the \textit{WMAP} power spectrum up to $l=200$ as the variance of the isotropic and Gaussian $a_{lm}$.
Although a sizable residual effect of the dipole can be seen in the maps, the entropies get modified only slightly. Clearly, at $l=1$ the entropies show the residual part of the dipole, and also at higher $l$ the curves are distorted a little, but the described procedure does not impose any large anomalies and especially it does not result in conspicuous values at $l=5,16,17,28,30$. Thus, we conclude that with our simplified approach no sizable mixing of dipole power to higher multipoles via the masking process can be observed. Finally note that other masking processes with Fourier methods were also applied to the angular and projection pseudoentropies in \cite{Mittelstaedt} and more extensively in \cite{OtgonbaatarMaster}\cite{Otgonbaatar}.

\subsection{Comparison of 2015 and 2018 data with angular pseudoentropy}

In the following we compare the angular pseudoentropy of 2015 \textit{Planck} data to the newest 2018 data release.

\begin{figure}
\begin{subfigure}[c]{\columnwidth}
\includegraphics[width = \textwidth]{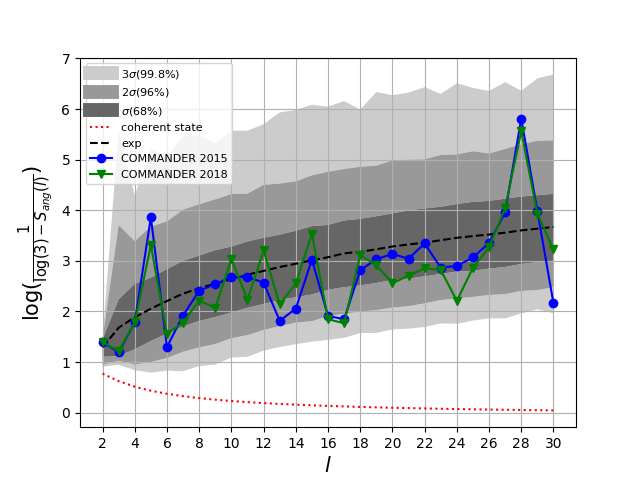}
\subcaption{Range $[2,30]$}
\end{subfigure}
\begin{subfigure}[c]{\columnwidth}
\includegraphics[width = \textwidth]{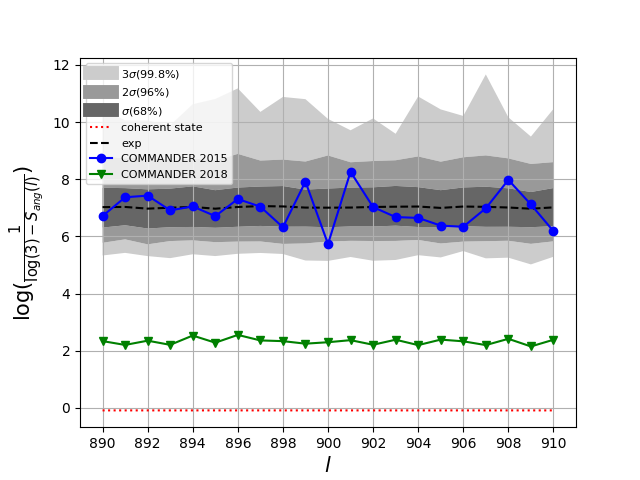}
\subcaption{Range $[890,910]$}
\end{subfigure}
\caption{Comparison of angular entropy with 2015 and 2018 data for COMMANDER.}
\label{15vs18_COMMANDER}
\end{figure}

\begin{figure}
\begin{subfigure}[c]{\columnwidth}
\includegraphics[width = \textwidth]{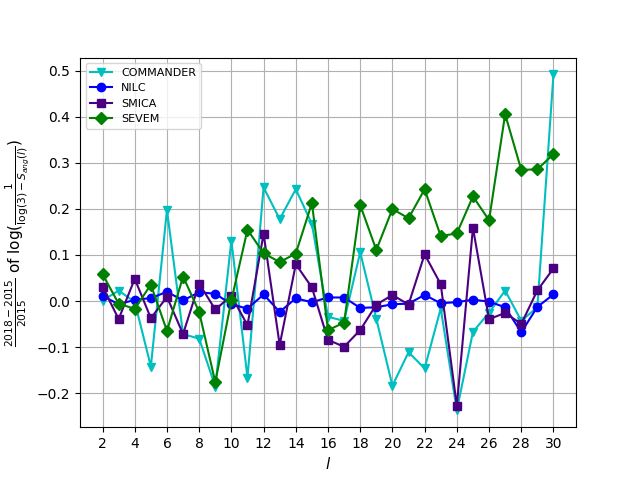}
\subcaption{Range $[2,30]$}
\end{subfigure}
\begin{subfigure}[c]{\columnwidth}
\includegraphics[width = \textwidth]{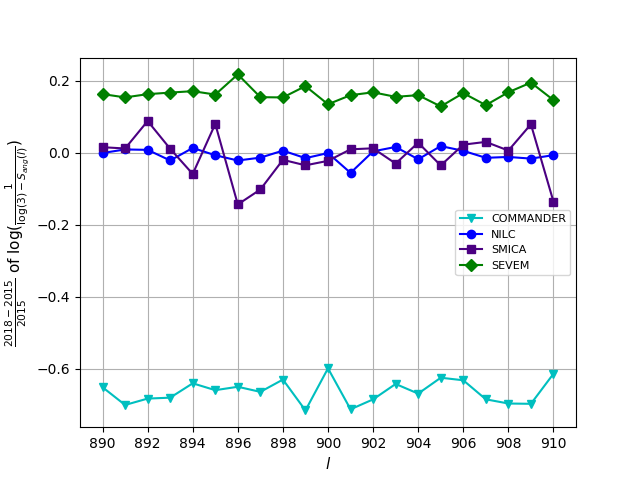}
\subcaption{Range $[890,910]$}
\end{subfigure}
\caption{Relative deviation of (logarithmic reciprocal deviation from maximal value of the) angular entropy with 2018 data compared to 2015 data for all maps.}
\label{All15vs18}
\end{figure}

Since the 2018 component separation process has been optimized for polarization data, it is expected to come equipped with a few drawbacks in temperature maps, especially for COMMANDER \cite{Planck2018IV}, which carries more residual foreground contamination in the 2018 than in the 2015 temperature map. One should expect to see this feature in the angular pseudoentropy and indeed Fig.~\ref{15vs18_COMMANDER}, which shows the comparison of 2015 and 2018 COMMANDER angular entropy on the two ranges considered in Sect.~\ref{results_ang}, as well as Fig.~\ref{All15vs18}, which shows the relative deviation of 2018 to 2015 data for all \textit{Planck} foreground cleaned full sky maps, confirm this expectation. While even for large angular scales the deviation of COMMANDER is larger than that of SMICA and NILC, for small angular scales COMMANDER drops even below SEVEM. Since in our work we do not want to mask the maps, but need to work with full sky data, it becomes obvious that for our purposes the 2015 COMMANDER temperature data should be preferred to the 2018 data.

\begin{figure}
\begin{subfigure}[c]{\columnwidth}
\includegraphics[width = \textwidth]{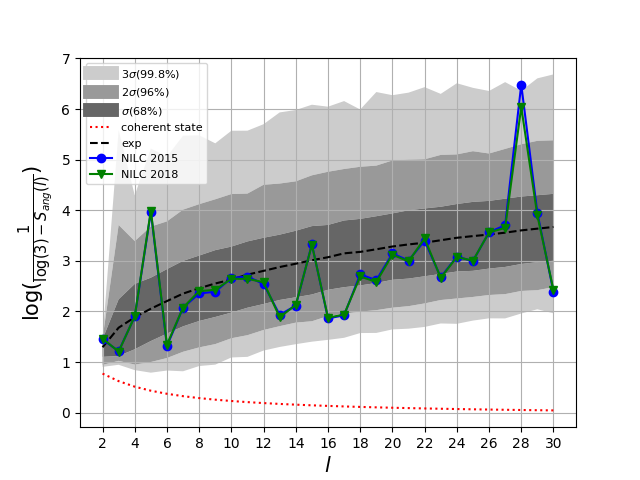}
\subcaption{Range $[2,30]$}
\end{subfigure}
\begin{subfigure}[c]{\columnwidth}
\includegraphics[width = \textwidth]{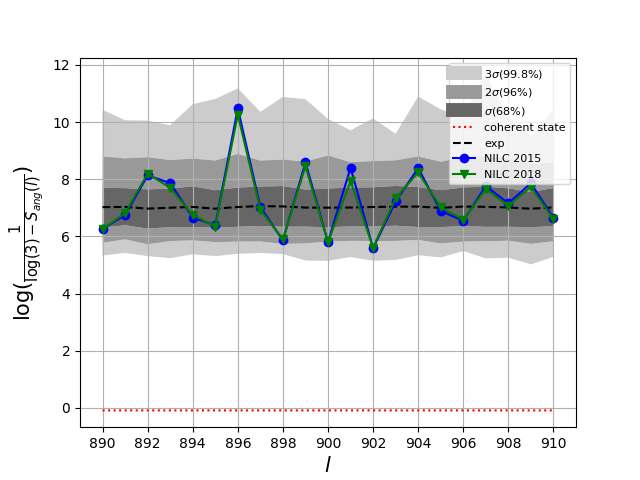}
\subcaption{Range $[890,910]$}
\end{subfigure}
\caption{Comparison of angular entropy with 2015 and 2018 data for NILC.}
\label{15vs18_NILC}
\end{figure}

\begin{figure}
\begin{subfigure}[c]{\columnwidth}
\includegraphics[width = \textwidth]{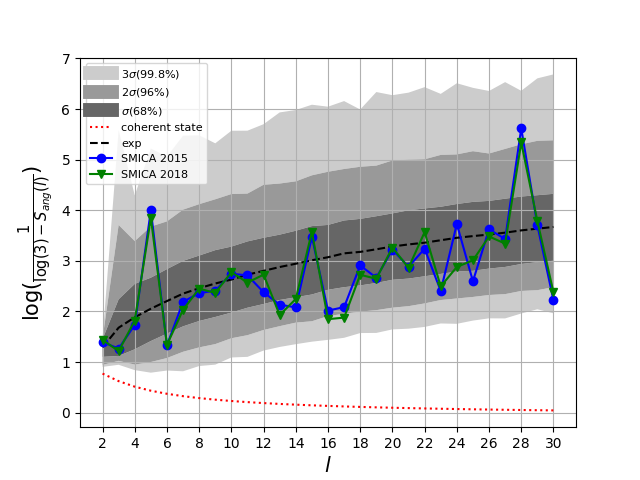}
\subcaption{Range $[2,30]$}
\end{subfigure}
\begin{subfigure}[c]{\columnwidth}
\includegraphics[width = \textwidth]{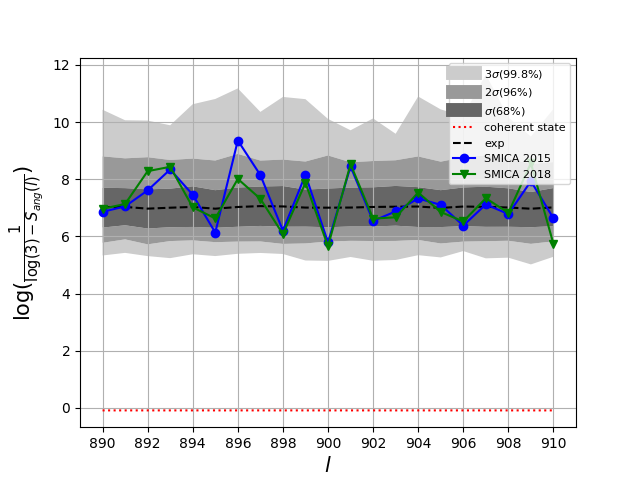}
\subcaption{Range $[890,910]$}
\end{subfigure}
\caption{Comparison of angular entropy with 2015 and 2018 data for SMICA.}
\label{15vs18_SMICA}
\end{figure}

Both NILC and SMICA show only a slight deviation in 2018 compared to 2015, both on small and large angular scales with NILC 2018 entropy being identical to the 2015 entropy with at most $5\%$ deviation, which is reached only at $l=28$, see Figs.~\ref{15vs18_NILC}, \ref{15vs18_SMICA} and \ref{All15vs18}. For all other multipoles the NILC deviation is nearly negligible. Since the NILC component separation process has been left nearly unaltered from 2015 to 2018, NILC is most useful for observing the effects of the improved Cosmic Dipole calculation and the removed AD non-linearity. The influence of the former is restricted mainly to a very slight reduction of significance of the both most unlikely multipoles we considered, namely $l=28,896$.

\begin{figure}
\begin{subfigure}[c]{\columnwidth}
\includegraphics[width = \textwidth]{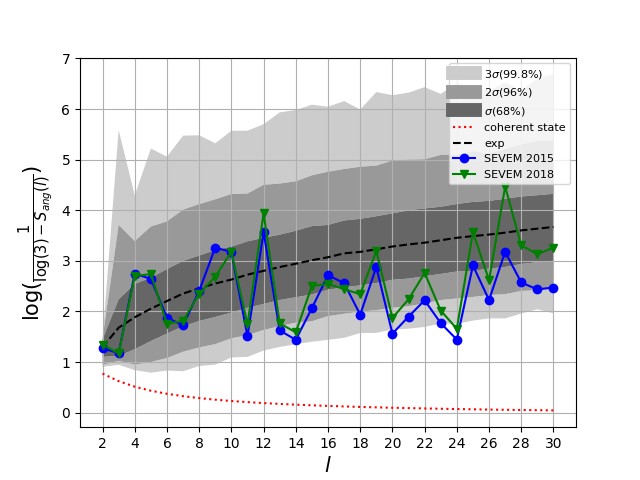}
\subcaption{Range $[2,30]$}
\end{subfigure}
\begin{subfigure}[c]{\columnwidth}
\includegraphics[width = \textwidth]{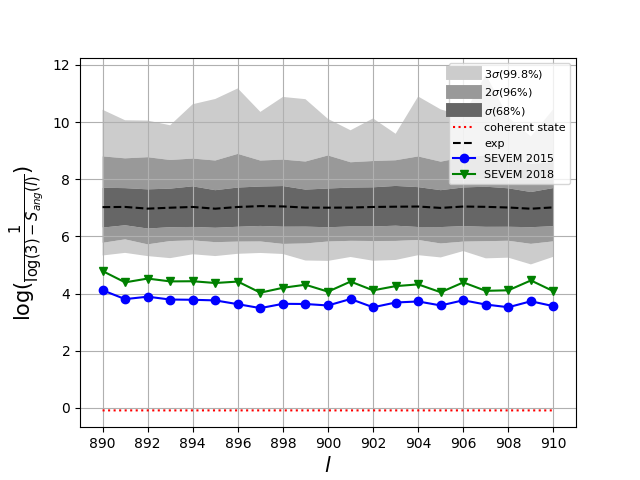}
\subcaption{Range $[890,910]$}
\end{subfigure}
\caption{Comparison of angular entropy with 2015 and 2018 data for SEVEM.}
\label{15vs18_SEVEM}
\end{figure}

SEVEM has been clearly enhanced in 2018, as shown in Figs.~\ref{15vs18_SEVEM} and \ref{All15vs18}. In 2015 data the angular entropy of SEVEM was far too low from $l=13$. That behavior came particularly clear at small angular scales. In the preceding sections we argued that this effect stems from the residual contamination of SEVEM data by the galactic plane. In 2018 the entropy is constantly shifted to higher values from $l=13$ on, approaching a nearly constant relative improvement of about $18\%$ at small angular scales. Nevertheless for our purposes the SEVEM map still lacks quality at small angular scales and visible residual foreground pollution is left.

\begin{figure}
\begin{subfigure}[c]{\columnwidth}
\includegraphics[width = \textwidth]{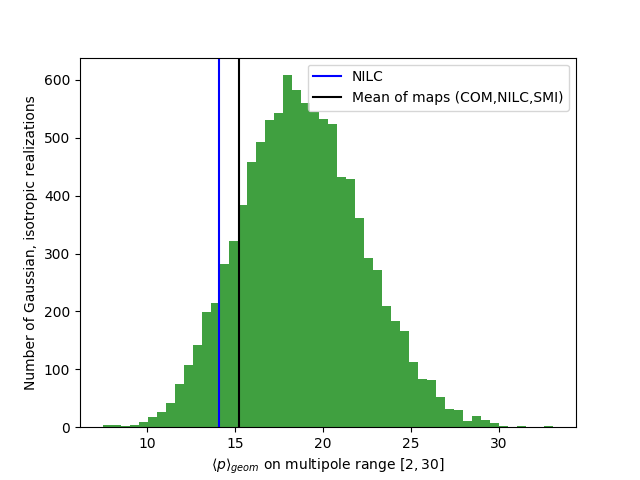}
\subcaption{Range $[2,30]$ calculated with 10000 ensembles.}
\end{subfigure}
\begin{subfigure}[c]{\columnwidth}
\includegraphics[width = \textwidth]{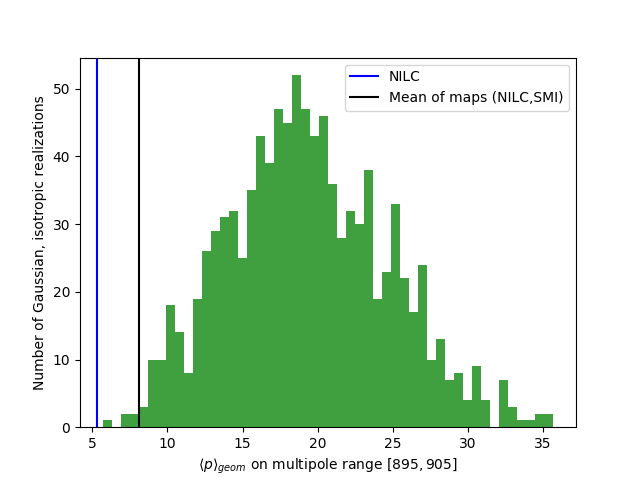}
\subcaption{Range $[895,905]$ calculated with 1000 ensembles.}
\end{subfigure}
\caption{Probability density of geometric mean of p-values and respective values for NILC and the geometric mean of maps for 2018 data.}
\label{Geommean2018}
\end{figure}

At large angular scales COMMANDER and SMICA exhibit a joint deviation behavior at the most unlikely multipoles. Both large entropy values at $l=5$ and $l=28$ are slightly suppressed, but still outside of $2\sigma$, and the small entropy value at $l=30$ is enlarged. While for SMICA all three multipoles still lie outside of $2\sigma$, COMMANDER shifts them towards smaller confidence. In contrast to that, the two multipoles $l=16,17$ get shifted to more unlikely values in both maps. These considerations show that the most conspicuous multipoles at large angular scales could partly be caused by unoptimized component separation, but that AD non-linearity and the Cosmic Dipole identification show only a minor effect, since NILC is nearly unaltered.

At small angular scales we can only use NILC and SMICA since SEVEM is still off and COMMANDER has been degraded due to polarization optimization. The main observation concerning single multipoles, that can be made at this point, is that for SMICA the most unlikely multipole $l=896$ on the considered range is improved but the previously normal multipole $l=910$ is shifted towards $2\sigma$.

In Tab.~\ref{tab_geommeans} in App.~\ref{app_plots} we gather p-values and likelihoods for 2015 and 2018 data. There we also include the range $[890,910]$ in order to compare it to $[895,905]$. Since COMMANDER is off at small angular scales in 2018, we also consider the geometric mean of NILC and SMICA alone. The range $[2,30]$ is normal in both data releases with 2015 being slightly better than 2018 both for NILC alone and for the geometric mean of COMMANDER, NILC and SMICA. Fig.~\ref{Geommean2018} shows that even though the geometric mean of p-values is smaller than the expectation, the significance for that is too low and hence we can conclude that using the angular entropy method the whole range $[2,30]$ is compatible with the assumption of isotropic and Gaussian temperature fluctuations, which was also pointed out in \cite{PinkwartSchwarz} for the range $[2,50]$.

In contrast the range $[895,905]$ displays unlikely behavior in both releases. We observe a slight enhancement from $4.4 \%$ to $6.2 \%$ in the geometric mean of p-values for NILC and from $5.3\%$ to $8.1\%$ in the geometric mean of p-values for the geometric mean of NILC and SMICA. These enhancements correspond to changes in likelihood from $0.1\%$ in 2015 to $0.5\%$ in 2018 for the latter and no change of likelihood for the former. None of thousand random isotropic and Gaussian maps admits such a low geometric mean of p-values as NILC in both 2015 and 2018, hence the Likelihood is bounded from above by $\approx 0.1\%$ in both releases. Enlarging the range a bit from $[895,905]$ to $[890,910]$ increases the likelihoods about a factor of $1$ to $15$, but keeps them under $2.5\%$. We can conclude that the improvements made in the 2018 data processing improve also the small angular scales, but the features are still at nearly $2\sigma$ for the mean of NILC and SMICA at $[890,910]$ and at or outside of $3\sigma$ for NILC at $[895,905]$. At this point we should clarify again that the question, how likely it is to find a range of such width outside of $3\sigma$ is postponed to the future and that here we might fall for the selection effect.

\subsection{Results for range angular pseudoentropy with 2015 data}

The range angular pseudoentropy provides an additional measure for quantifying unlikeliness of multipole ranges and also collections of different multipoles which are not necessarily in a row, see Eqs.~\eqref{psi_range}-\eqref{ang_range} in Sect.~\ref{math}. Unfortunately it is a mix of a correlation and an averaging measure, hence one needs to consider both the range angular and the single multipole angular entropies in order to identify effects of correlation of different multipoles, which are usually expressed by small range entropies. On the other hand, if one is solely interested in the mean likelihood of a given range, the geometric mean of p-values of the single multipole angular entropy is surely the better measure. Aside the partial correlation interpretation, the big advantage of the range angular entropy is its pseudoentropy nature and henceforth its interpretation as an entanglement measure. In Tab.~\ref{pvaltab} in Appendix \ref{app_plots} we gather p-values for the range angular entropy for various ranges on large angular scales and for the small scale range $[895,905]$, as well as the signed deviation of the entropy for 2015 NILC from the isotropic and Gaussian expectation.

One directly observes that the range entropy of the range $[2,3]$ is too small at $2\sigma$. Having in mind Fig.~\ref{ang30} this low p-value should mainly stem from correlation of $l=2$ and $l=3$. Since the single value angular entropy is related to multipole vectors via their chordal distances, we propose that this feature is the same as the correlation of the quadrupole and octupole multipole vectors (see \cite{Schwarz2015} for more details).

The next observation concerns the range $[2,6]$. Even though $l=5$ is too high at $\approx 1\%$ p-value and $l=6$ too low at $\approx 7\%$ p-value, the only unlikely collection of multipoles on this range containing $l=5$ or $l=6$ is the collection $\{2,5\}$ which gives a range entropy that is too small at nearly $2\sigma$-level. Since the average of $\{2,5\}$ should not differ significantly from $\{3,5\}$, which in turn has a p-value of above $50\%$, the low p-value of $\{2,5\}$ stems from anti-correlation or entanglement of $l=2$ with $l=5$. The multipole vectors of $l=2$ and $l=5$ are unusually widespread over the sphere. Hence, we draw the conclusion that the unusually large value of $l=5$ concerning the single value angular entropy is induced by the CMB quadrupole, which itself is mainly influenced by the Cosmic Dipole, that is assumed to constitute the main ingredient of large scale multipole vector anomalies (see \cite{PinkwartSchwarz}). We furthermore point out that to the authors knowledge this (anti-)correlation of the CMB quadrupole with $l=5$ has not been observed so far.

The range $[2,30]$ yields a p-value of around $18\%$ which is compatible with the likelihood of about $15\%$ calculated with the geometric mean of p-values for the single multipole angular entropy. The range entropy lies slightly below the expectation indicating a mixture of a slight averaged preference for a direction on the sky and a slight correlation of multipoles, though being within $1\sigma$.

For the collection $\{5,28\}$ the p-value is smaller than one would expect from the average of both multipoles indicating correlation of these two multipoles.

On small angular scales the p-value $0.5\%$ for the range $[895,905]$ is compatible with the likelihood from the geometric mean, which we gave the approximate upper bound $0.1\%$. For both ranges $[2,30]$ and $[895,905]$ we obtain slightly larger p-values with the range entropy than likelihoods with the geometric mean, which could be caused by reduction of significance due to averaging of large and small values of the angular entropy. The fact that the range entropy lies below the expectation again indicates, that a direction might be preferred in the data and/or different multipoles might be correlated.

\section{Summary and Discussion}
\label{sum}
Building upon the Wehrl entropy we introduced three types of pseudoentropies which approximate the Wehrl entropy but allow for much faster computation and hence analyzing CMB data up to $l=1000$. Those entropies are the $j$-projection, the quadratic and the angular entropy. While the quadratic is simple in fashion it should be disregarded, because of its $x^2$ instead of $x\log(x)$ behavior and the numerical problems one runs into with it for high $l$ if not approximating the expression. All pseudoentropies are rotationally invariant measure of quantum randomness or entanglement on spin-$l$ states and hence on each multipole of CMB temperature fluctuations on the sphere. Contrary to the usual von Neumann entropy these pseudoentropies do not vanish for pure states. In the spirit of thermodynamics, the entropies are useful for reducing $2l+1$ d.o.f. per multipole to a single number per multipole just as one usually does with $C_l$ but which complements it in the case of anisotropies or non-Gaussianities.

We showed that for $l=2$ both the Wehrl and the angular entropy depend only on the squared chordal distance of multipole vectors, yielding another view on this method and a connection to many previous studies of CMB analysis.

Although our focus was on introducing the methods and clarifying their properties, in order to demonstrate the usage of these methods we applied the introduced types of pseudoentropies for analysis of CMB temperature full sky maps and it turned out that they all show similar behavior and the same characteristic features of the maps, except for the quadratic pseudoentropy. Since the angular entropy is the computationally cheapest measure, the rest of the analysis was devoted solely to the angular entropy, which reaches its maximum $\log(3)$ for maximally mixed states, which cannot be reached by pure temperature maps, and its minimum probably for coherent states; it is mathematically known for sure for $l=1/2,1$ only. The physical data from the \textit{Planck} 2015 maps and \textit{WMAP} ILC was compared to isotropic and Gaussian maps, and some multipoles with particularly small p-values were found, in particular $l=5,16,17,28,30$ on large angular scales, and the range $895 \leq l \leq 905$ for NILC, the likelihood of which we approximately bounded from above by $0.1\%$, but there could be as well further unusual angular scales and we did not take into account the selection effect statistically. On average three out of four \textit{Planck} maps do not show abnormal global behavior for $l \leq 1000$, that means deviations from isotropy and Gaussianity on a large range of scales, and the abnormality of the fourth map -- SEVEM -- can be removed by masking the galactic plane. As expected, the \textit{Planck} maps can be considered as a clear improvement compared to \textit{WMAP} on small angular scales $l>200$. 

A comparison of isotropic, Gaussian random maps to maps constructed from uniform multipole vectors showed that our method is sensitive to deviations from Gaussianity, resp.~statistical dependence of spherical harmonic coefficients, and, due to its rotationally invariant nature, also isotropy. One should note that uniform multipole vectors are clearly distinguishable from isotropic and Gaussian maps for which the multipole vectors exhibit repulsion.

We were not able to identify the masking process as a reason for all or some of the mentioned conspicuous multipoles with a simple masking approach. The fact that these multipoles have low p-values in all of the maps except for SEVEM indicates a different reason behind them. Nevertheless, one should not withhold that a statistical fluke cannot be excluded, even if the p-value for $l=28$ lies below half a percent. 

Comparing \textit{Planck} 2015 to 2018 data confirmed the expectation that the COMMANDER full sky 2018 map cannot be used for our methods at large angular scales without masking. The SEVEM full sky map has been enhanced from 2015 to 2018 but still carries too much foreground pollution in it when not masked. SMICA only deviates slightly and NILC is nearly unaltered. The unlikely features we observe are present in both data releases but with slightly less significance in 2018 than in 2015. The fact that NILC is left nearly unchanged suggests that AD non-linearity and unoptimized Cosmic Dipole removal do not account for the observed features. Nevertheless the component separation still might do. 

Eventually we considered the angular range entropy as a mixture of a measure of range-or-collection-averaged angular entropy and of correlation between different multipoles. It turned out that the results for the ranges $[2,30]$ and $[895,905]$ are consistent with the likelihoods obtained from the geometric mean of p-values of the standard angular entropy. We found the anti-correlation of the quadrupole with $l=5$ and the correlation of $l=2,3$ especially interesting. While the latter supports the previously observed quadrupole-octupole correlation, the former hints towards a connection between the high angular entropy value at $l=5$ and the CMB dipole, which itself has been proposed to be influenced mainly by the Cosmic Dipole in the past.

There are several tasks left for the future. First of all a deeper statistical analysis needs to be done, especially considering wider and smaller angular scale ranges. Our analysis has shown, that there might be something hidden at small angular scales and it would be interesting to see more results on this. For identifying possible foreground effects it would be useful to apply our methods to the foreground maps from \textit{Planck} data and investigate to what extent certain features are foreground residuals of the component separation. Furthermore the influence of noise, especially inhomogeneous noise, on the results needs to be evaluated. So far we have introduced the methods, explained some of their mathematical behavior and performed a perfunctory analysis in order to illustrate their usage. Furthermore one could try to apply some of these methods to polarization data. For the Wehrl entropy it is clear that a direct generalization is possible, but for the angular entropy further analysis is needed. Angular and Wehrl entropy are also useful in quantum information theory as real entanglement measures. Finally, our work is also interesting from a mathematical perspective. The fact that the Wehrl entropy is minimized in general by SU(N)-coherent states is known as the Lieb conjecture and has been proven \cite{Lieb16}. The same question is, however, still open for the angular entropy.

\begin{acknowledgments}
Based on observations obtained with \textit{Planck} (http://www.esa.int/\textit{Planck}), an ESA science mission with instruments and contributions directly funded by ESA Member States, NASA, and Canada; and \textit{WMAP} (https://map.gsfc.nasa.gov/).
 
We acknowledge financial support by the Deutsche Forschungs Gemeinschaft Research Training Group 1620 \emph{Models of Gravity}. We thank Valentin Buciumas, Jessica Fintzen,  Otgonbaatar Myagmar, and Gregor Mittelstaedt for collaboration at earlier stages of the research leading to this publication. Useful discussions with Elliott Lieb are gratefully acknowledged. We also thank the referee for useful comments and suggestions.
\end{acknowledgments}

\bibliographystyle{cj}
\bibliography{mybiblio} 


\clearpage

\appendix

\section{System resources}
\label{system}
System resources used include the following:
\begin{itemize}
\item[(i)] Model: OptiPlex 790, Dell Inc., \\ Version 01, 64 bits;
\item[(ii)] CPU: IntelCore i5-2400, $3.10 \, \mathrm{GHz}$, 1 CPU, \\ 4 Cores, 4 Threads;
\item[(iii)] Cache: L1 -- $256 \, \mathrm{kB}$, L2 -- $1 \, \mathrm{MB}$, L3 -- $6 \, \mathrm{MB}$;
\item[(iv)] RAM: $4 \, \mathrm{GB}$;
\item[(v)] GFlops (tested with linpack): \\From $50.9$ up to $76.2$
\end{itemize}
\vfill

\section{Additional plots and tables}
\label{app_plots}

\begin{figure}[ht]
\includegraphics[width=\columnwidth]{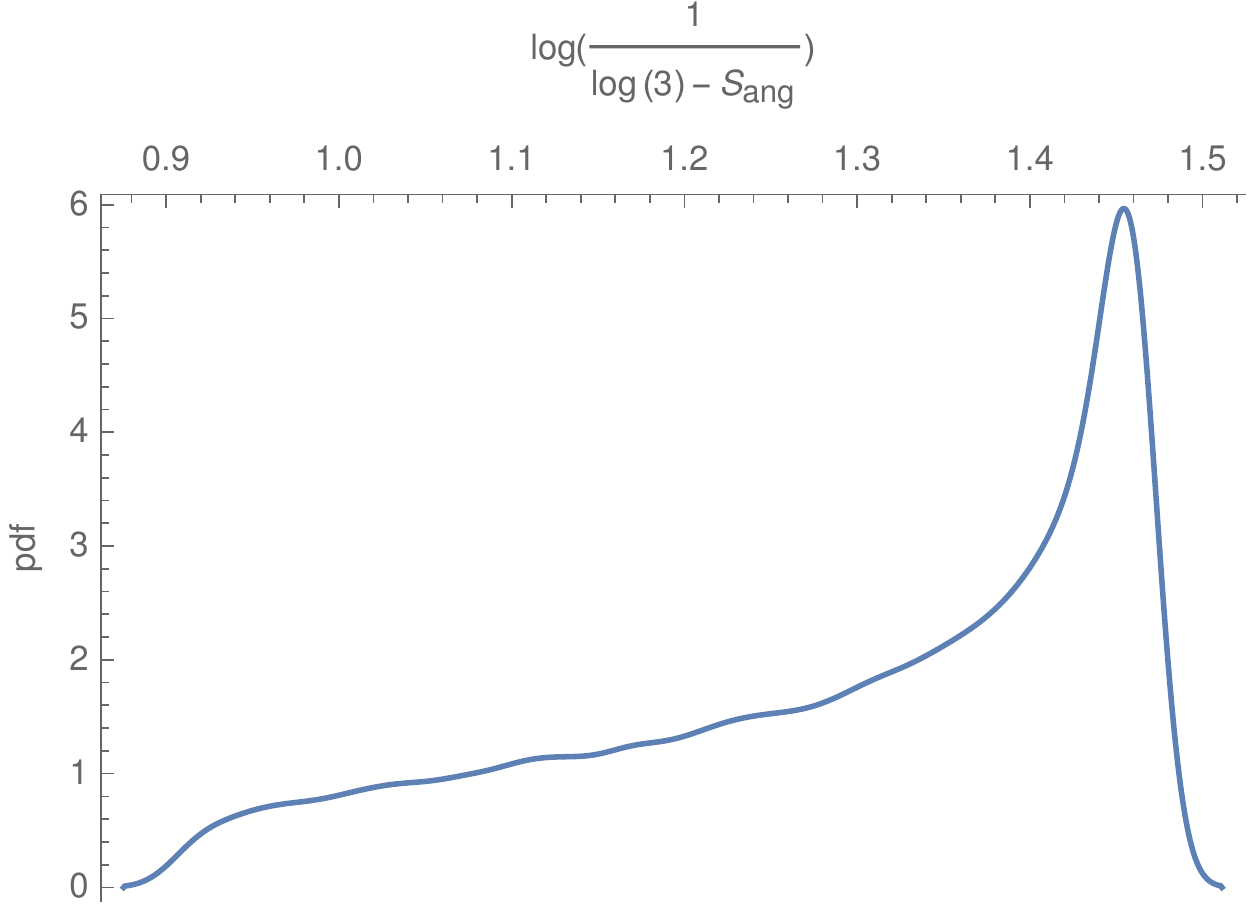}
\caption{Probability distribution of the logarithmic reciprocal of the angular pseudoentropy for Gaussian, isotropic $a_{lm}$ $l=2$; calculated with $10^5$ random ensembles and smoothed.}
\label{l2Gauss}
\end{figure}

\begin{figure}
\includegraphics[width=\columnwidth]{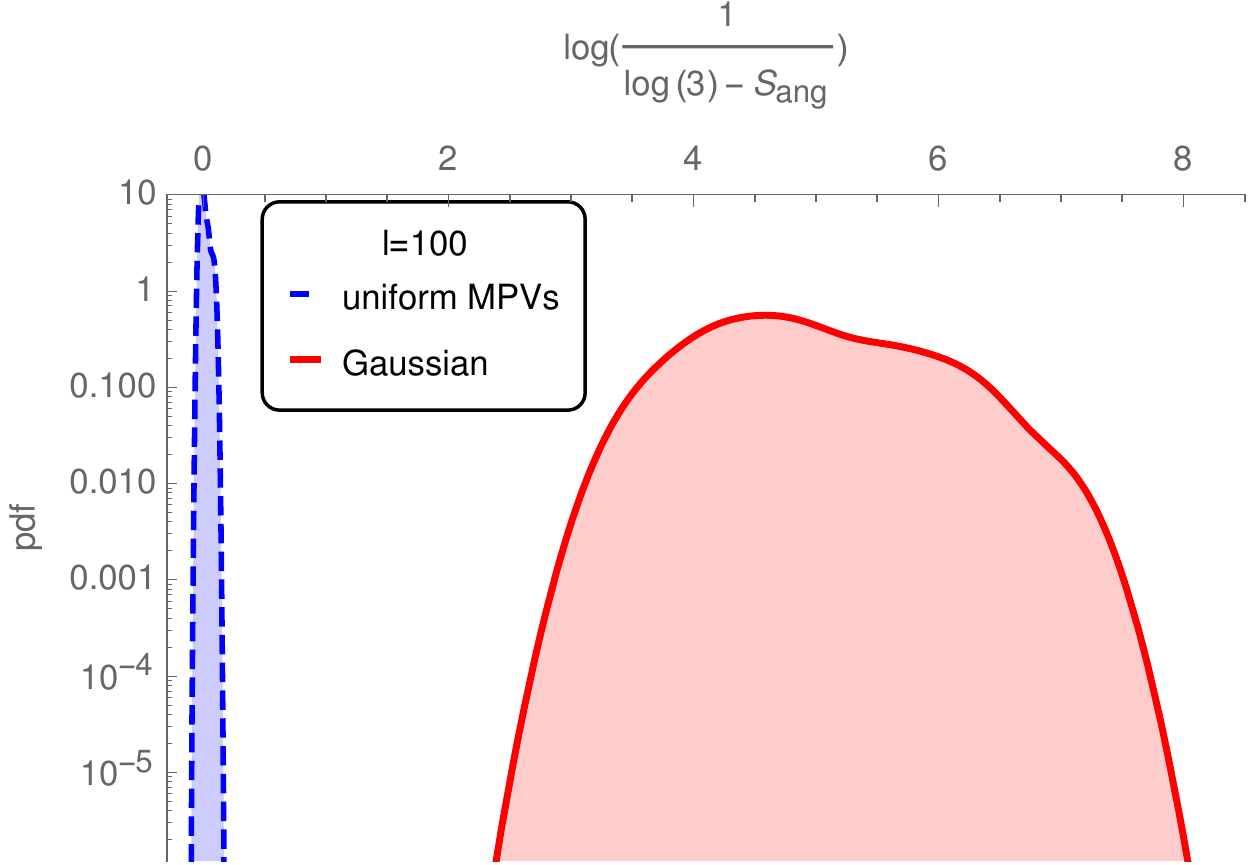}
\caption{Comparison of the probability distribution of the logarithmic reciprocal of the angular pseudoentropy between uniform MPVs and isotropic, Gaussian $a_{lm}$ at $l=100$; calculated with $10^2$ random ensembles and smoothed.}
\label{l100}
\end{figure}

\begin{figure}
\includegraphics[width=\columnwidth]{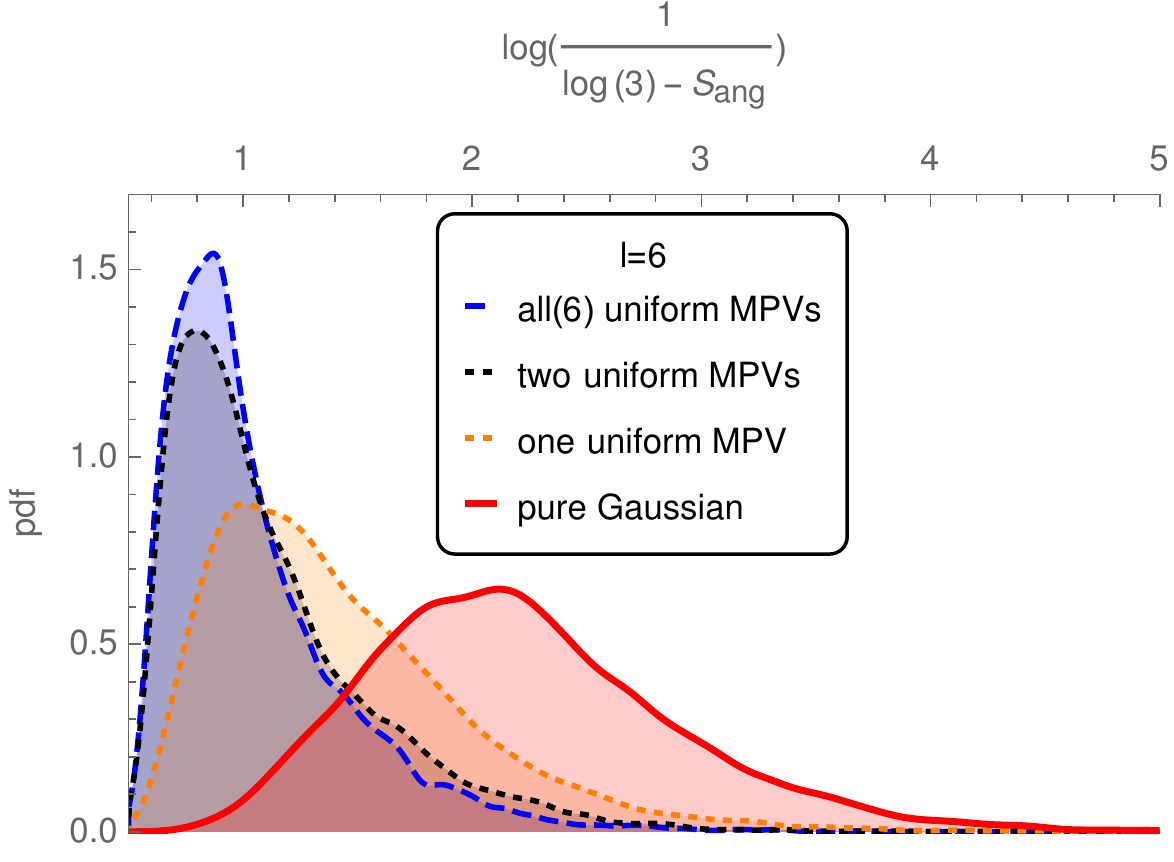}
\caption{Comparison of the probability distribution of the logarithmic reciprocal of the angular pseudoentropy of maps -- constructed from $n=0,1,2,6$ uniform MPVs and $m=6-n$ MPVs extracted from isotropic and Gaussian maps -- at $l=6$; calculated with $10^4$ random ensembles and smoothed.}
\label{GaussAndUni}
\end{figure}

\begin{figure}
\includegraphics[width = \columnwidth]{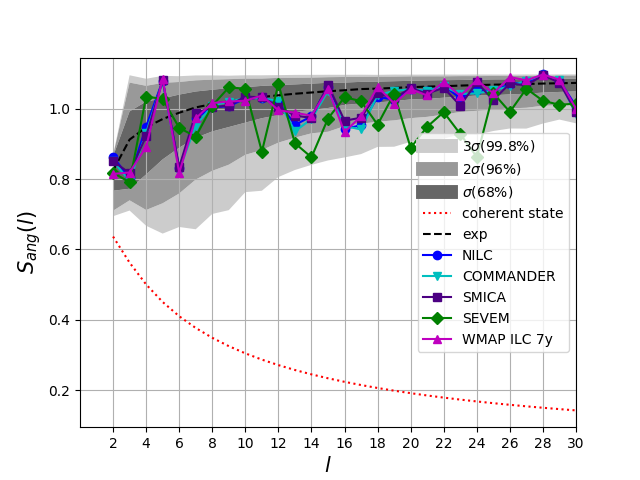}
\caption{Angular pseudoentropy plotted unlogarithmicly between $l=1$ and $l=30$. Confidence levels have been smoothed.}
\label{unlog_ang}
\end{figure}

\begin{figure}
\includegraphics[width = \columnwidth]{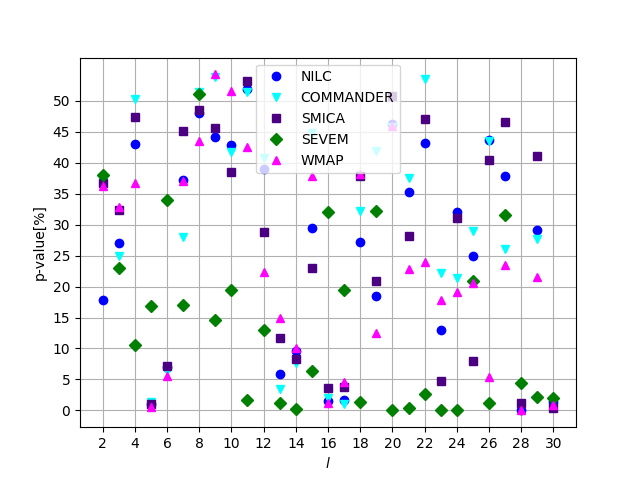}
\caption{P-values of angular pseudoentropy between $l=1$ and $l=35$, calculated with 10000 ensembles of Gaussian $a_{lm}$}
\label{pvals_plot}
\end{figure}

\begin{figure*}
\begin{subfigure}[c]{0.49\textwidth}
\includegraphics[width = \textwidth]{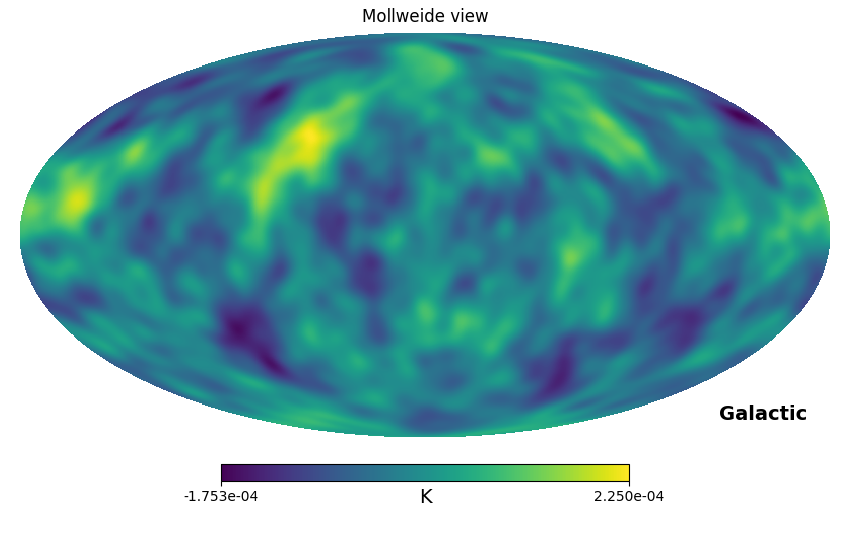}
\subcaption{Map from isotropic and Gaussian $a_{lm}$ with power spectrum (up to $l=200$) of \textit{WMAP} 7-year ILC, smoothed}
\end{subfigure}
\begin{subfigure}[c]{0.49\textwidth}
\includegraphics[width = \textwidth]{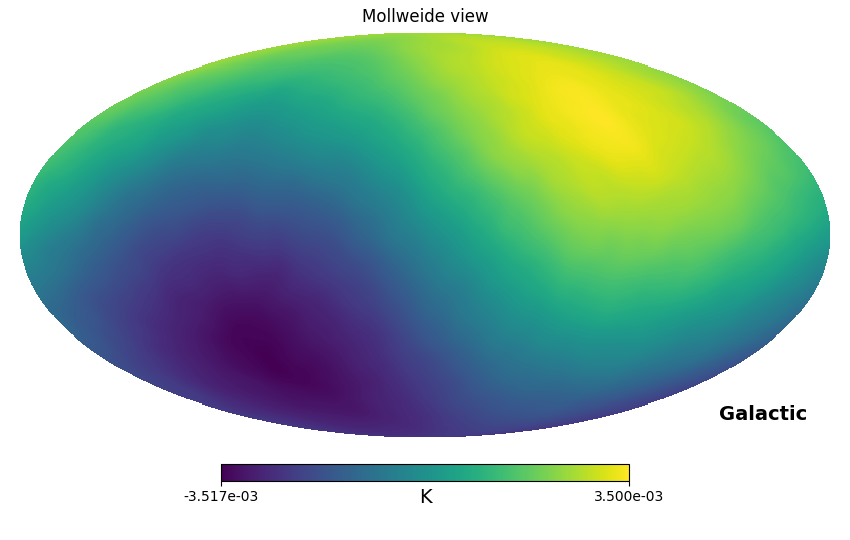}
\subcaption{Map after induction of dipole}
\end{subfigure}
\begin{subfigure}[c]{0.49\textwidth}
\includegraphics[width = \textwidth]{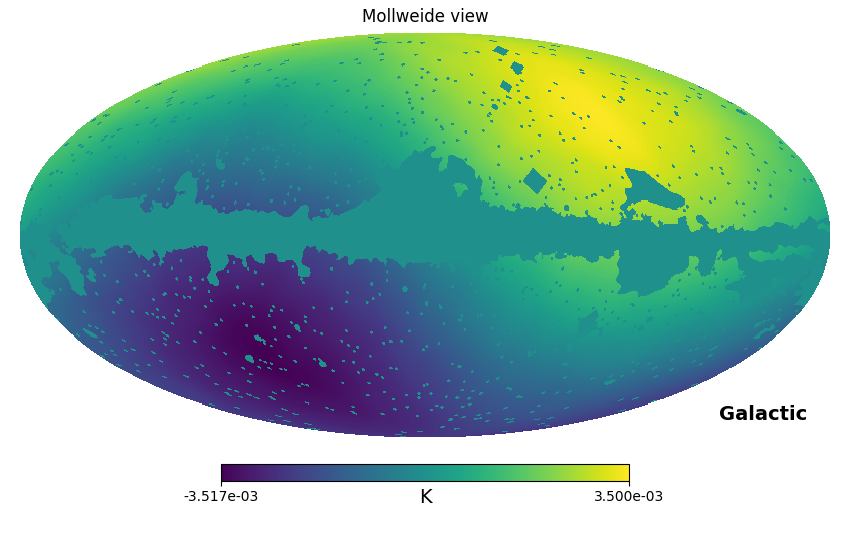}
\subcaption{Map after induction of dipole and masking with \textit{WMAP} intensity mask}
\end{subfigure}
\begin{subfigure}[c]{0.49\textwidth}
\includegraphics[width = \textwidth]{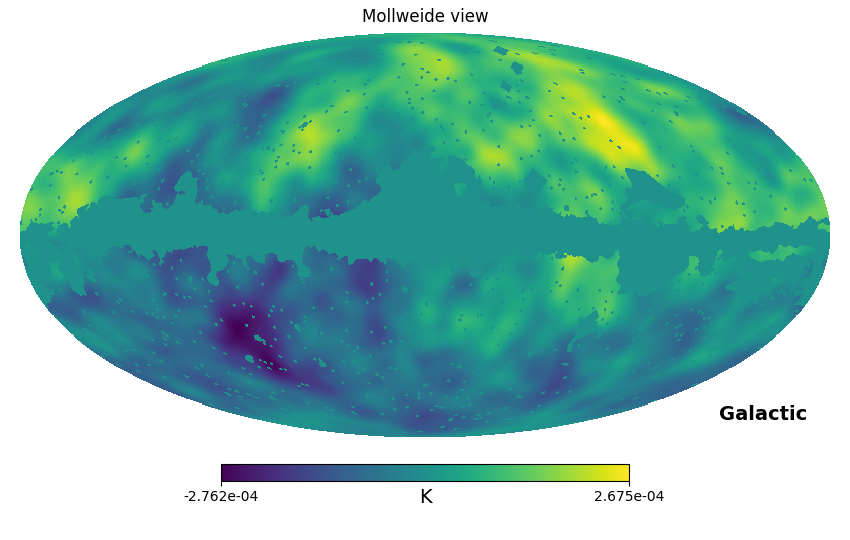}
\subcaption{Map after removing dipole again with Healpy remove\_{}dipole}
\end{subfigure}
\begin{subfigure}[c]{0.49\textwidth}
\includegraphics[width = \textwidth]{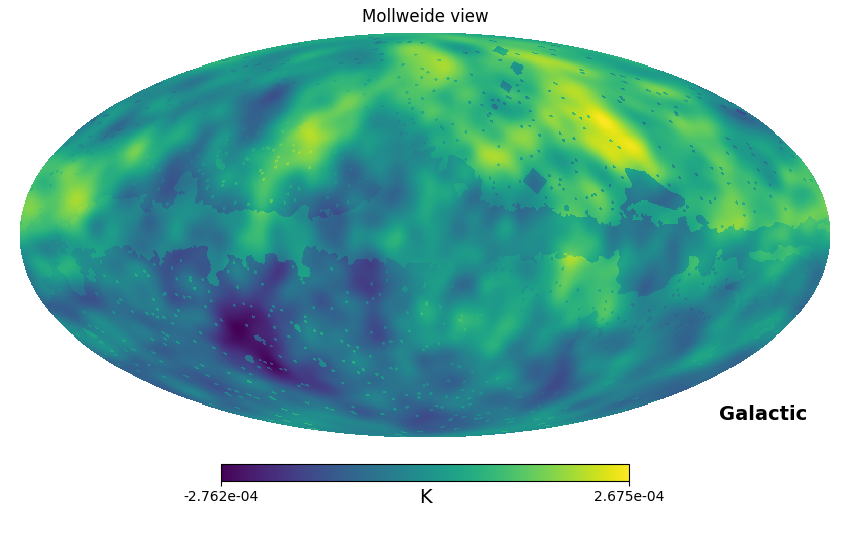}
\subcaption{Map after removing dipole and filling masked region with data from a)}
\end{subfigure}
\caption{Process of inducing a dipole, masking the map, removing the dipole and filling the masked region.}
\label{maps}
\end{figure*}

Figures \ref{l2Gauss},\ref{l100},\ref{GaussAndUni},\ref{unlog_ang},\ref{pvals_plot},\ref{maps} and Tables \ref{Fit_params_ang},\ref{tab_geommeans},\ref{pvaltab} are given in this Appendix.

\begin{figure}
\begin{subfigure}[c]{\columnwidth}
\includegraphics[width = \columnwidth]{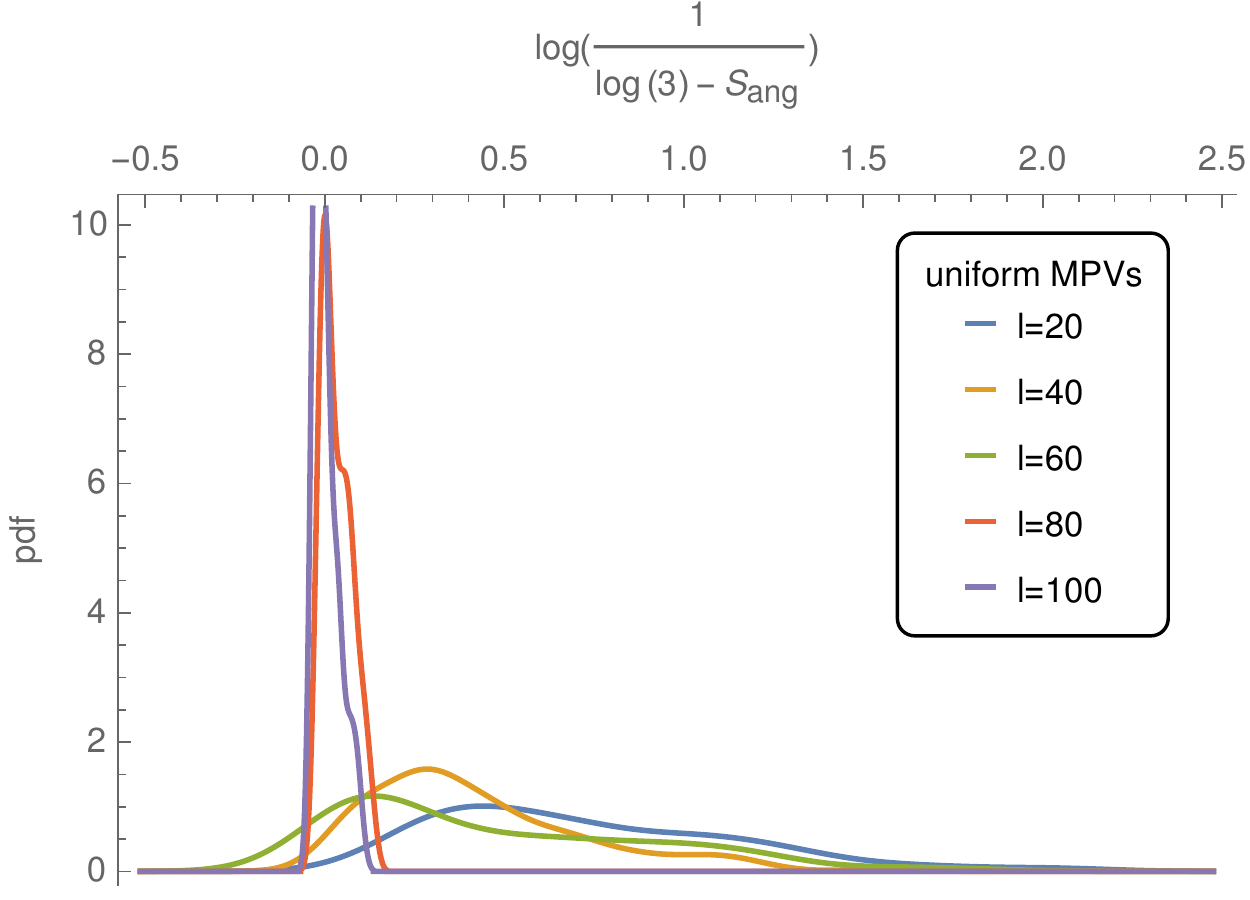}
\end{subfigure}
\begin{subfigure}[c]{\columnwidth}
\includegraphics[width = \columnwidth]{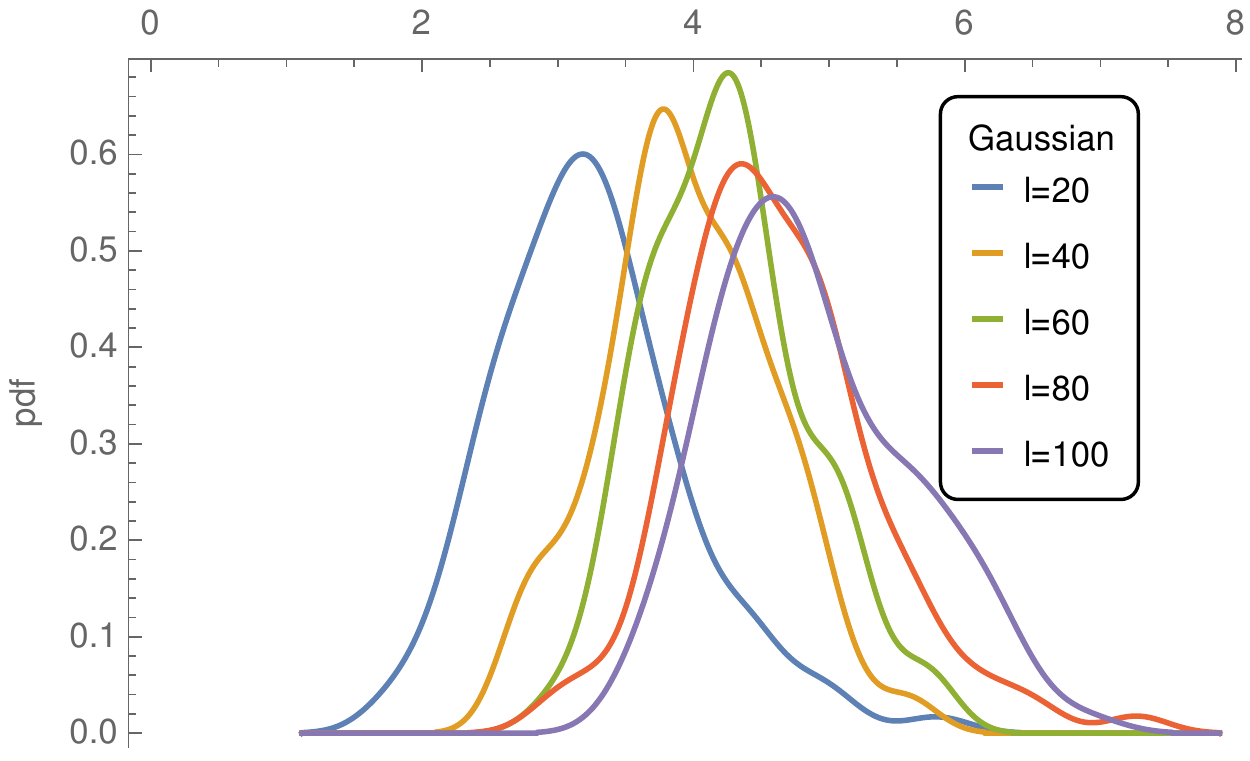}
\end{subfigure}
\caption{Probability distribution of the logarithmic reciprocal of the angular pseudoentropy using maps from uniform MPVs (top) and Gaussian, isotropic $a_{lm}$ (bottom) at $l=20,40,60,80,100$; ; calculated with $10^2$ random ensembles and smoothed.}
\label{prob_highl}
\end{figure}

\begin{table}[ht]
\begin{tabular}{|c||c|c|c|}
\hline
Quantity & a & b & c \\
\hline \hline
$1\sigma_{\mathrm{uppper}}$ & 0.658 & 0.999 & $\cdots$ \\
\hline
$1\sigma_{\mathrm{lower}}$ & 0.663 & 0.929 & $\cdots$ \\
\hline
$2\sigma_{\mathrm{upper}}$ & 1.672 & 1.742 & $\cdots$ \\
\hline
$2\sigma_{\mathrm{lower}}$ & 1.198 & 0.477 & $\cdots$ \\
\hline
$3\sigma_{\mathrm{upper}}$ & 2.982 & 68.283 & $\cdots$ \\
\hline
$3\sigma_{\mathrm{lower}}$ & 1.654 & 0.345 & $\cdots$ \\
\hline
Expt. & 0.948 & 1.545 & 0.976 \\
\hline
\end{tabular}
\caption{Fitted parameters of angular pseudoentropy rounded to three decimals.}
\label{Fit_params_ang}
\end{table}

\begin{table*}
\begin{tabular}{|c||c||c|c|c|}
\hline
Range & Year & Map & $\langle p \rangle_{\mathrm{geom}}$(\%) & Likelihood $L$(\%) \\
\hline \hline
\multirow{2}{*}{[2,30]} & \multirow{2}{*}{2015} & NILC & 14.68 & 11.42 \\
& & Mean of COM, NILC, SMI & 15.67 & 18.29 \\
\cline{2-5}
& \multirow{2}{*}{2018} & NILC & 14.07 & 8.20 \\
& & Mean of COM, NILC, SMI & 15.22 & 14.89 \\
\hline
\multirow{2}{*}{[895,905]} & \multirow{3}{*}{2015} & NILC & 4.4 & $\lessapprox$ 0.1 \\
& & Mean of COM, NILC, SMI & 8.6 & 0.8 \\
& & Mean of NILC, SMI & 6.2 & 0.1 \\
\cline{2-5}
& \multirow{3}{*}{2018} & NILC & 5.3 & $\lessapprox $ 0.1 \\
& & Mean of COM, NILC, SMI & $\lessapprox $ 0.1 & $\lessapprox $ 0.1 \\
& & Mean of NILC, SMI & 8.1 & 0.5 \\
\hline
\multirow{2}{*}{[890,910]} & \multirow{3}{*}{2015} & NILC & 8.8 & $\lessapprox $ 0.1 \\
& & Mean of COM, NILC, SMI & 13.6& 8.2 \\
& & Mean of NILC, SMI & 10.8 & 1.5 \\
\cline{2-5}
& \multirow{3}{*}{2018} & NILC & 10.3 & 0.7 \\
& & Mean of COM, NILC, SMI & $\lessapprox$ 0.1 & $\lessapprox$ 0.1 \\
& & Mean of NILC, SMI & 11.5 & 2.4 \\
\hline
\end{tabular}
\caption{Comparison between 2015 and 2018 of geometric mean of p-values and likelihoods for all three ranges calculated for NILC, the geometric mean of COMMANDER, NILC and SMICA and the geometric mean of only NILC and SMICA. For large angular scales $l\in [2,30]$ we used $10^4$ ensembles and two digits for the p-value and likelihood; for small angular scales $\l \in [895,905]$ and $l \in [890,910]$ we used $10^3$ ensembles and one digit for the p-value and likelihood.}
\label{tab_geommeans}
\end{table*}

\begin{table*}
\begin{center}
\begin{tabular}{|c|c|c|c|c|}
\hline
Range & p-value $[\%]$ & Entropy & Expectation & Above or below  \\
\hline \hline
$[2,3]$ & 2.0 & 0.820 & 0.960 & - \\
$[2,4]$ & 44.3 & 1.015 & 1.016 & - \\
$[2,5]$ & 41.4 & 1.057 & 1.045 & + \\
$[2,6]$ & 55.9 & 1.055 & 1.054 & + \\
\hline 
$[3,4]$ & 27.5 & 1.030 & 0.991 & + \\
$[3,5]$ & 20.9 & 1.066 & 1.035 & + \\
$[3,6]$ & 40.2 & 1.060 & 1.047 & + \\
\hline
$[4,5]$ & 11.8 & 1.067 & 1.015 & + \\
$[4,6]$ & 36.0 & 1.056 & 1.038 & + \\
\hline
$[5,6]$ & 20.3 & 1.051 & 1.007 & + \\
\hline
$\{2,5\}$ & 3.2 & 1.077 & 0.998 & + \\
$\{3,5\}$ & 51.6 & 1.013 & 1.008 & + \\
\hline
$\{4,6\}$ & 30.6 & 0.990 & 1.007 & - \\
\hline
$5$ & 1.03 & 1.080 & 0.988 & + \\
$6$ & 6.95 & 0.831 & 1.004 & - \\
\hline
$[2,20]$ & 17.9 & 1.0847 & 1.0857 & - \\
$[2,30]$ & 18.4 & 1.0789 & 1.0893 & - \\
\hline
$[26,27]$ & 40.8 & 1.085 & 1.080 & + \\
$[26,28]$ & 16.6 & 1.093 & 1.086 & + \\
$[26,29]$ & 24.7 & 1.094 & 1.090 & + \\
$[26,30]$ & 19.9 & 1.088 & 1.092 & - \\
\hline
$[27,28]$ & 6.4 & 1.094 & 1.081 & + \\
$[27,29]$ & 12.2 & 1.094 & 1.087 & + \\
$[27,30]$ & 17.3 & 1.086 & 1.090 & - \\
\hline
$[28,29]$ & 3.9 & 1.095 & 1.082 & + \\
$[28,30]$ & 23.7 & 1.084 & 1.088 & - \\
\hline
$[29,30]$ & 4.4 & 1.062 & 1.083 & - \\
\hline
$\{5,28\}$ & 2.3 & 1.084 & 0.986 & + \\
\hline
$28$ & 0.07 & 1.097 & 1.071 & + \\
\hline
$[895,905]$ & 0.5 & 1.09827 & 1.09851 & - \\
\hline
\end{tabular}
\caption{Range angular pseudoentropy: The second column shows p-values for different multipole ranges and collections, the third and fourth columns show the value of the range angular pseudoentropy for NILC and the expectation using 10000 ensembles of isotropic and Gaussian random maps, the fifth column indicates if the NILC entropy lies below and above the expectation, which is important for the interpretation of the results. The single multipole angular entropy values for $l=5,6$ are included for comparison with the ranges that include both multipoles.
}
\label{pvaltab}
\end{center}
\end{table*}

\vfill

\end{document}